%% file: ms9.tex
\documentclass[useAMS,usenatbib,usegraphicx]{mn2e}
\usepackage{verbatim}
\title[Low z \lyaf\ simulations]{The \lyaf\ at redshifts 0.1 -- 1.6:
good agreement between a large hydrodynamic simulation and HST spectra}

\author[Paschos \etal]{Pascal Paschos\thanks{E-mail: ppaschos@minbari.ucsd.edu},
  Tridivesh Jena, 
  David Tytler,
  David Kirkman\\
  \newauthor and Michael L. Norman \\
  Center for Astrophysics and Space Sciences,
  University of California San Diego,
  La Jolla, CA, 92093-0424 \newauthor
}

\input{astmac}

\begin{document}

\date{\today}

\maketitle

\begin{abstract}

We give a comprehensive statistical description of the \lya\ absorption
from the intergalactic medium in a hydrodynamic simulation at redshifts 
0.1 -- 1.6, the range
of redshifts covered by spectra of QSOs obtained with the Hubble Space
Telescope. We use the ENZO code to make a simulation in 
a cube with side length 76~Mpc comoving using 75~kpc cells, for a Hubble
constant of 71~\kms Mpc$^{-1}$. The best prior work, by \citet{dave99},
used an SPH simulation in a 15.6~Mpc box with an effective resolution of
245~kpc and slightly different cosmological parameters. We use a popular
cosmological model and astrophysical parameters
that describe the ultraviolet background that photoionizes and heats the
gas. At redshifts $z =2$ this simulation is different from data.
\citet{tytler07b}
found that the simulated spectra at $z=2$ have too little power on large 
scales, \lya\ lines are too wide, there is a lack high
column density lines, and there is a lack of pixels with low flux. 
Here we present statistics at $z < 1.6$, including the flux
distribution, the mean flux, the effective opacity, and the power and
correlation of the flux. We also give statistics of the \lya\ lines
including the line width distribution, the column density distribution,
the number of lines per unit equivalent width and redshift, and the
correlation between the line width and column density. We find that the
mean amount of absorption in the simulated spectra changes smoothly with
redshift with DA($z$) $=0.0102(1+z)^{2.252}$. Both the trend and absolute 
values are close to measurements of HST spectra by \citet{kirkman07a}.
 The column density distribution and the line width distribution are also
 close to those measured from HST spectra by \citet{janknecht06a},
except for the mode of the line width distribution which is smaller
in the HST spectra, probably because of photon noise. Although some differences
 that we saw at $z=2$ are too subtle to be seen in existing HST spectra, 
overall, the simulation gives a good description of HST spectra at 
$0.1<z<1.6$.

\end{abstract}


\section{Introduction}

We are exploring the physical conditions in the IGM using hydrodynamic
numerical simulations. Our goal is to make simulations that give simulated
H~I \lya\ absorption that is statistically identical to that measured in
spectra of QSOs, where the best data give errors on the statistics of a
few percent. We hope that such simulations will provide a comprehensive,
accurate and reliable description of the physical conditions in the IGM.

In previous papers we compared our optically thin ENZO simulations to data
at redshifts $z \sim 2$.  In \citet{jena05a}(J05) we found that
popular values of cosmological and astrophysical parameters gave simulated
spectra that simultaneously match both the
the mean amount of \lya\ absorption and the distribution of \lya\ 
line widths in ground based spectra of QSOs.
In \citet{tytler07b} however, we found that the same simulation does not
simultaneously match other statistics measured in data. The
power spectrum of the flux in the simulated spectra has significantly
lower amplitude than we measure in data, by 50\% on large
scales and 20\% on small scales. The lines in the simulated
spectra are wider with larger $b$-values, and the simulated
spectra have different distributions for both the flux per pixel and the
number of lines at various column densities than do the data. Here we use the
same simulation to explore redshifts 0.1 -- 1.6 where the \lya\ line is
visible in spectra obtained from space.

\citet{dave99} used SPH simulations to explore the
\lyaf\ at these low redshifts. One of their simulations used cosmological
parameters close to those favoured today: $\Omega_m = 0.4$, $\Omega_b =
0.0473$, $\Omega_\Lambda = 0.6$, $h = 0.65$ and $\sigma_8 = 0.8$. However,
they used a box only 15.6~Mpc (we use a Hubble constant of 71~\kms Mpc$^{-1}$
throughout the paper)
on a side and an effective resolution of
245~kpc, both far from ideal. It is now well known
\citep{pen97a,barkana04a,sirko07a,tytler07b}, that we need box sizes of
the order of a hundred Mpc to contain the long mode density perturbations.
In smaller boxes, we lack these long wavelength modes and we see
significantly lower velocities and temperatures at all densities including
those that make the \lyaf\
\citep{tytler07b}. We also require small enough cells that we can
effectively resolve the structures in the \lyaf . \cite{meiksin04a} found
that the ideal cell size is $< 40$~kpc and in \citet{tytler07b} we confirm
this. The 75~kpc cells that we use here are large enough that the line
widths are significantly larger than they would be with much smaller
cells. J05 showed that a cell size of $150$~kpc gives a poor
representation of the IGM, sometimes showing trends with input parameters that
are the opposite of what we see with small cells that resolve the \lyaf .

In \S \ref{sec:secenzo}
below, we briefly describe the simulation code and parameters we have
adopted. In \S \ref{sec:fluxstat} we describe the statistics of the flux
in the \lyaf\ including the mean flux, flux distribution and the flux
power and autocorrelation function. In \S \ref{sec:linestat} we describe the
statistics of the lines in the \lyaf\ including the $b$-values, column densities
and their correlations. In \S \ref{sec:datacomp} we give a brief
comparison to data from HST. We discuss and conclude in \S \ref{sec:conc}.

\section{ENZO IGM Simulations}
\label{sec:secenzo}

The numerical simulation that we describe in this paper
\citep{bodenheimer07a} uses the Eulerian hydrodynamic cosmological code
ENZO
\citep{bryan95a,bryan97a,norman99,oshea04a,oshea05a,regan07a,norman07a}.
The simulation contains both CDM and baryons in the form of gas, but no
stars. The simulation was run with popular values for cosmological parameters:
a flat geometry $\Omega_{total}$ = 1, comprising a vacuum energy
density of $\Omega_{\Lambda}=0.73$, $\Omega_{m}=0.27$ (CDM plus baryons),
a baryon density of $\Omega_{b}=0.044$, a Hubble constant of $H_0$ = 71~km
s$^{-1}$ Mpc$^{-1}$ and an initial power spectrum scalar slope of $n_s =
1.0$ with a current amplitude of \sig $=0.9$.

\begin{figure}
\includegraphics[width=84mm]{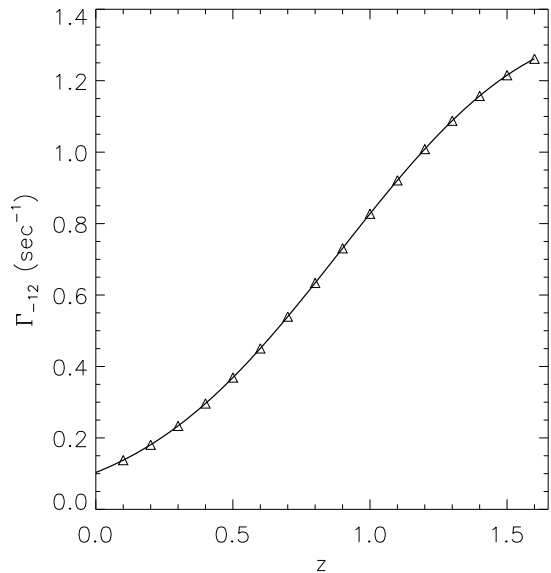}
\caption{The photoionization rate from the UV emitted by QSOs and galaxies
calculated by \citet{haardt01a} that we use in this simulation. The
vertical axis is in units of $10^{-12}$
photoionization per neutral H atom per second.
}
\label{figgamma12}
\end{figure}

The ENZO code follows the evolution of the gas using non-equilibrium
chemistry and cooling for hydrogen and helium ions
\citep{abel97a,anninos97a}. After reionization at $z=6$, photoionization
is provided using the \citet{haardt01a} volume average UV background (UVB)
from an evolving population of galaxies and QSOs.  We show the rate of
ionization of neutral hydrogen in 
Fig.~\ref{figgamma12}

The simulations are optically thin so that all cells experience the same
UV intensity at a given time. We do not treat the transfer of radiation
inside the volume, and we include no feedback from individual stars, galaxies 
or QSOs except for that implied by the uniform UVB.

As in J05, we use two astrophysical
parameters to describe the intensity of the UVB. The
parameter \gammah\ is the rate of ionization per H~I atom in units of the
Haardt \& Madau model discussed above, while \gammahe\ measures the heat
input per He~II ionization, again in units of the rate for the Haardt \&
Madau spectrum.
In J05 we showed how we can use \gammahe\ to adjust the widths of the
\lya\ lines and the mean amount of absorption in simulated spectra,
while \gammah\ will also adjust the mean amount of absorption.
We can then choose values for \gammahe\ and \gammah\ that together
match both the line widths and mean flux in data.
The simulation that we discussed in this paper uses \gammah\ = 1.0 and
\gammahe\ = 1.8, and we called it simulation or box A in
\citet{jena05a}(J05) and \citet{tytler07b}.

We initiate the simulations using an \citet{eisenstein99} power spectrum
for the dark matter perturbations, that we insert at $z=60$. The simulated
volumes are all cubes with strictly periodic boundary conditions. Hence
the power is input at a finite number of discrete wavenumbers.

The amplitude of the power that we insert varies smoothly with wavenumber.
We insert the amplitude expected for the universe as a whole, with no
random variations associated with the finite box sizes. The power in the
simulations is not adjusted to include the variations in mean density that
we see in the universe on the scale of the boxes. In this limited sense,
the boxes contain information on scales much larger that their sizes.

The simulation has a comoving box size of 76.8~Mpc and a comoving cell
size of 53.25$h^{-1}$ or 75~kpc comoving. The simulation has one CDM
particle for each cell initially, and each dark matter particle has a mass
of $M_{CDM} \simeq 9.5 \times 10^{6}~h^{-1} M_{\odot}$.

Our simulation is barely large enough that the values we obtain at $z=0.1$
will be reliable. The nonlinear scale can be quantified by calculating the 
wavenumber $k_{NL}$ such that 
$\frac{1}{(2\pi)^3} \int_{0}^{k_{NL}} P^{3D}(k) 4 \pi k^2 dk = 1$ where 
$P^{3D}(k)$ is the three dimensional matter power spectrum. This indicates 
that the nonlinear scale is 14.11 Mpc (comoving) at $z=1.6$ and 25.51 Mpc
(comoving) at $z=0.1$. These correspond to 18.38\% of the box size at 
$z=1.6$ and 33.22\% of the box size at $z=0.1$.  
In smaller boxes, the largest modes that fit inside the
box have become non-linear before this low redshift and the results will
be unreliable.

\subsection{Comparison of this simulation to data at $z= 2$}

In \citet{jena05a}(J05) we showed that at redshift 1.95 various simulations
using typical cosmological and astrophysical parameters gave a good match
to both the mean flux transmitted in the \lyaf\ and the $b$-value
distribution that we use to describes the line widths. We noted in Fig. 40
of J05 that the
power spectrum of the flux for these simulations was similar to that of
data, but we did not examine this in detail.

In \citet{tytler07b} we compared a wider variety of statistics at $z=2$
to the simulations that we use in this paper,
 and to other simulations using the same
input parameters in smaller boxes.
We found that if we matched the mean flux
well, the line widths in the simulation were too big using $\sig = 0.9$
and \gammahe = 1.8 times the standard UVB heating.  If we instead use the smaller
\sig\ values preferred by the WMAP 3 year data, the lines in the simulated
spectra would be still larger. We now know that we 
could better match the line
widths using \gammahe\ $ < 1.8$, corresponding to a softer ionizing
spectrum that gives less heating. However, 
from Fig. 39 of J05, we would simultaneously
need to use \gammah $>1$ to
preserve the match with the mean absorption, and hence it is unclear
whether the combination of \gammahe\ and \gammah\ parameters would be 
concordant with other data.

We also found that the simulated spectra had too little power on both
small and large scales. The small scales are an alternative way of
noticing that the lines are too wide in the simulations. However, the
larger scales are harder to adjust. We might use a larger \sig\ value or
we might need to adjust the heating.

We found that our simulations have too many lines with \lnhi $< 14$ \cmm,
a slight lack of lines with \lnhi = 14--15 \cmm\ and a large lack with
\lnhi $> 17$ \cmm. We did not investigate whether these differences are
themselves enough to account for the missing power. We also found,
confirming \citet{bolton07d}, that our simulated spectra have too few
 pixels with a lot of absorption.

We concluded that there were three classes of changes that might allow us
to make simulations that exactly match data. 

First, there might be problems with the simulations. We use moderate
resolution optically thin simulations that we hope will match the low
density IGM.
\citet{tytler07b} showed that using a much larger box does not help,
while a factor of four smaller cell size made the difference larger.
\citet{kohler07a} showed that they obtain the correct number of Lyman limit
systems (LLS, with \lnhi $>17.2$~\cmm ) at
$z=4$ using 2~kpc cells with radiative transfer.  Perhaps a much larger
version of their simulation might also
 match the entire column density distribution, the power, line widths and
flux distribution. We should certainly add radiative transfer, especially
that associated with the ionization of He~II by discrete sources. This
should give more realistic temperatures and may improve the match to the
data \citep{bolton07d}. Early results such as \citet{bolton04a} and
\citet{paschos07} show that radiative transfer increases the temperature
of the IGM at the cosmic mean density, the opposite of what we need.
However radiative transfer might also give lower temperatures at the higher
densities that make the \lyaf\ \citep{bolton07d}.

A second option is to change the astrophysics and especially the heating
of the IGM. A softer UVB will probably make spectra that are more like
data. The third choice is to change the cosmological parameters, specifically
we expect that a simulation with \sig $>0.9$ would make spectra with more large
scale power, more like the data.

In summary, we know that our simulation is not ideal, but they are as
good as any yet done, and a major improvement on those published  at $z < 2$.

When we compare simulations to data we would like to concentrate on the
low density IGM that we can most readily simulate. We take this to mean
\lnhi $< 17$~\cmm\ since the lack of lines becomes much larger at higher
column densities and it is relatively easy to find Lyman continuum
absorption in real spectra. At $z=2$ the simulations contain some such lines, but
factors of many less than there are in real spectra \citep[Fig.
15]{tytler07b}. Many of the statistics that we show below include lines of all
column densities, including the few lines with \lnhi $> 17$~\cmm\ in our
simulated spectra. We will be analysing 2000 spectra at each redshift
from 0.1 to 1.6. These spectra contain no lines with \lnhi $>17$~\cmm\ at
$z=0.1$ and 26 such lines at $z=1.6$.

\section{Statistics of the Flux in the \lya\ Forest}
\label{sec:fluxstat}

We make  spectra of the flux transmitted through the IGM at \lya\
wavelengths using the code described in J05 and following the
improvements discussed in
\citet{tytler07b}. We made 2000 spectra, each parallel to $z$-axis of the
box and beginning at a random location on the $xy$-face of the box.  We
made 2000 spectra at each redshift from 0.1 -- 1.6, in steps of 0.1.  The
random numbers that selects the $x$, $y$ coordinates of each spectrum
on the cube's face were different at
each redshift and hence these spectra are approximately independent,
except that they all sample the same large scales structure in the box.

We use a number of pixels per spectrum equal to the number of cells along
a box side.  We calculate the optical depth as a function of velocity 
in a spectrum, $\tau(v) $, using equations in
\S 4 of J05, and the flux from $F = exp ( -\tau )$, where $F= 1.0$ in the
absence of absorption, and $F \rightarrow 0$ at the centre of saturated
absorption lines. The spectra from all simulations are smooth functions of
$v$, with no discreteness from the cell size, because to determine $\tau $
at each position we integrate over at least $\pm 512$ cells, to the left
and to the right of the absorbing cell. Each spectrum is made at
a fixed redshift. We convert from the Mpc per grid cell into
velocity $v$ in \kms\ of a spectrum using the $H(z)$ for the chosen $z$,
but we do not increment the $z$ as we move down a sight line, and we do
not change the H($z$). Each spectrum is then frozen at the indicated
redshift with no evolution.

\begin{figure*}
\includegraphics[width=84mm]{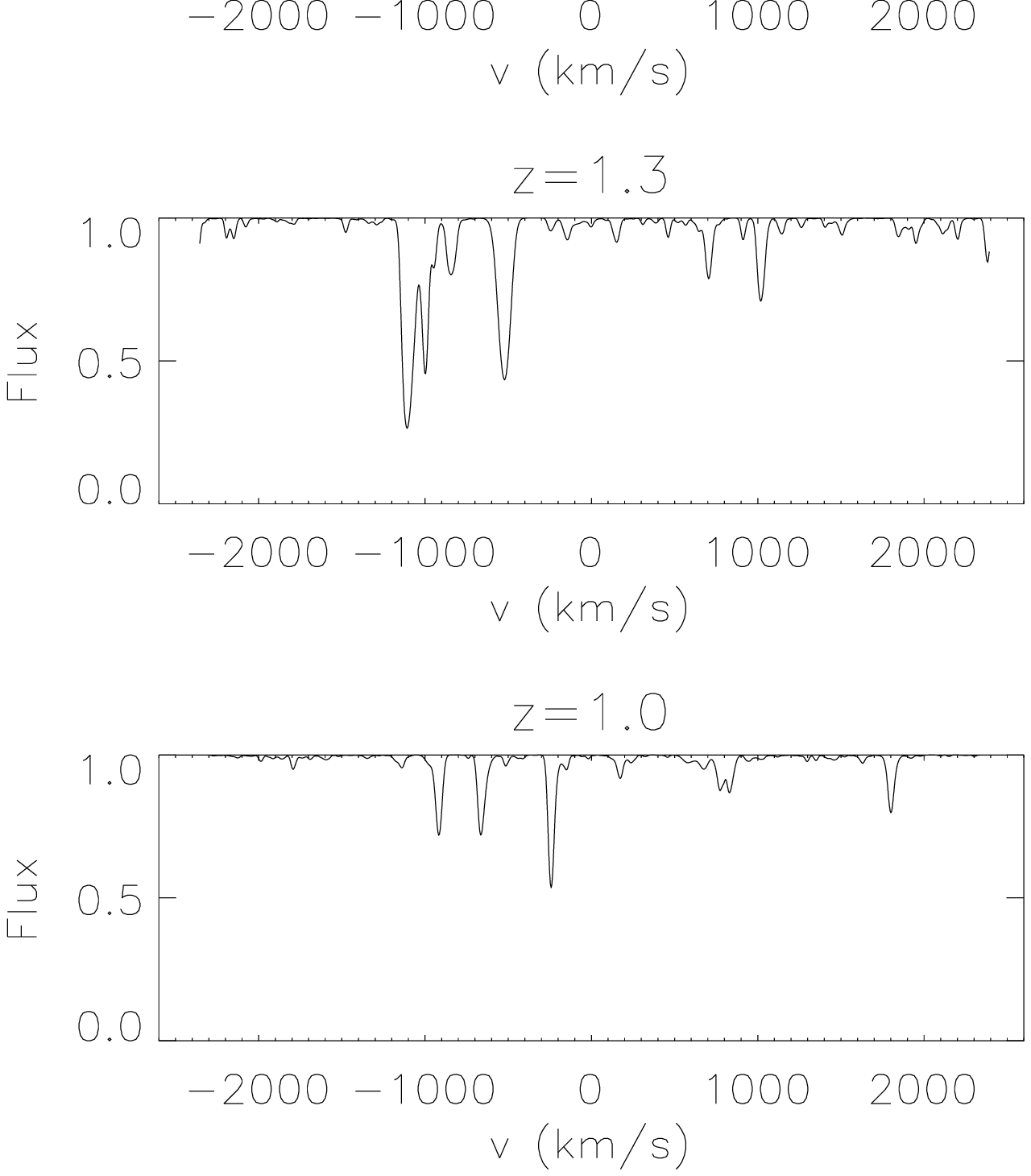}
\includegraphics[width=84mm]{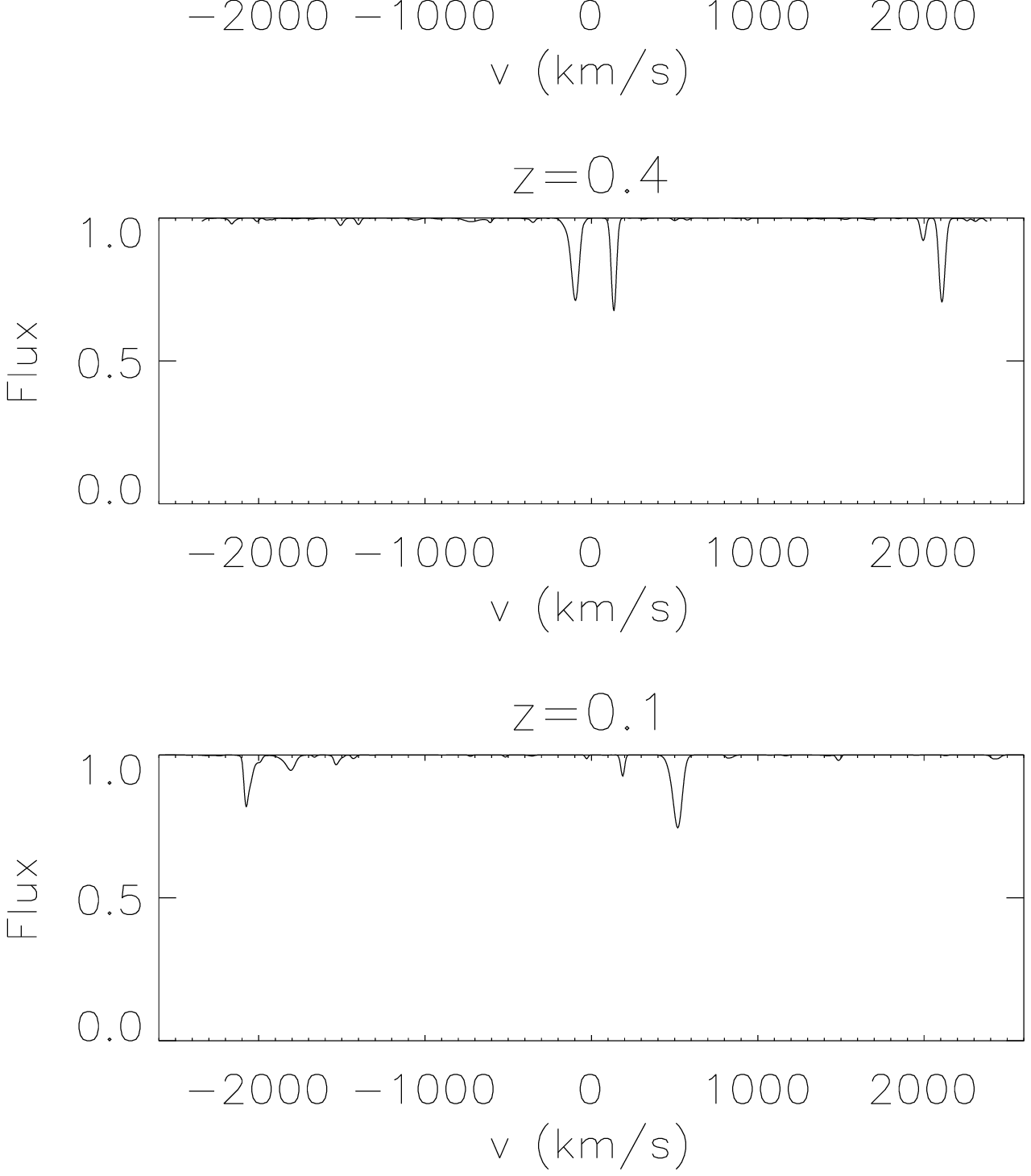}
\caption{Spectra of the \lya\ absorption from the simulation.
For each redshift we choose to show a spectrum with a mean flux that is
approximately  the same as the mean flux of all the spectra at that $z$.
Although the velocity span of the box, and hence the spectra,
decreases with the Hubble constant at low redshifts, we plot the same
velocity range for all spectra.
}
\label{spectra}
\end{figure*}

In Fig. \ref{spectra} we show examples of the spectra.
At lower redshifts we see systematically less absorption in total, fewer
strong lines with large equivalents, and more regions with less than 5\%
absorption.

\subsection{Effective Opacity}
\label{taueff}

\begin{figure}
\includegraphics[width=84mm]{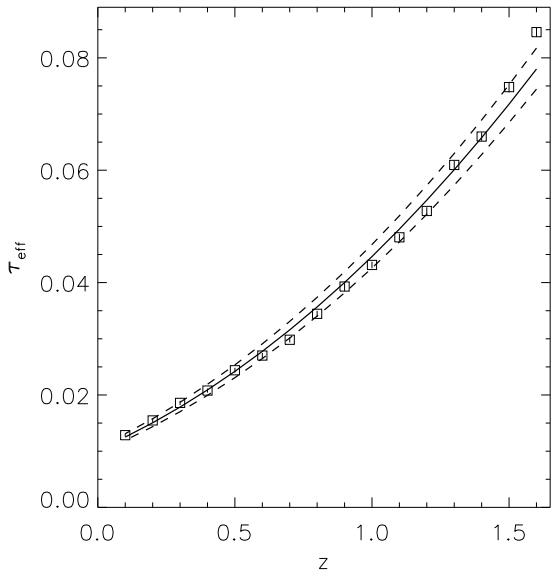}
\caption{The effective opacity and a power law fit. The fit is a function
of the form ${\tau_{eff}^{fit} = A (1+z)^\alpha}$ where $z$ is the redshift,
${A = 0.0102}$ and ${\alpha = 2.252}$. The curves above and below the solid
line are best fits to the 2$\sigma$ upper and lower limits on the
individual points. They do not show the range of possible fits.
}
\label{figtaueff}
\end{figure}

In Fig. \ref{figtaueff} we show the mean effective opacity as a function
of redshift. We calculate the effective opacity from the mean flux
${\bar{F}}$ using ${\tau_{eff} = -ln(\bar{F})}$, as is normally done. The
mean flux is the mean of all spectra that we made at each redshift. In
Table \ref{sim_table} we give values for this and other statistics that
have a single value at each redshift.

We see that the effective opacity fall smoothly with decreasing redshift,
from  0.085 at $z = 1.6$ to 0.013 at $z = 0.1$. This is
well fit by the power law that we show.

\begin{table}
\caption{\label{sim_table} Statistics of the \lyaf\ at redshifts 0.1
to 1.6. $\Gamma _{12}$ is the ionization rate in units of $10^{-12}$
ionizations per H atom per second. $\tau _{eff}$ is the mean effective
optical depth, and DA the mean absorption. The values in the
err column are $10^4$ times the standard deviation of the DA values.
Under $\sigma ^2 _L $ we give $10^3$ times the variance in the mean flux
in each spectrum.
For $\sigma ^2$ we list 100 times the variance in
the mean flux in each pixel in all the spectra
$F_{mode}$ is the mode of the mean flux in each spectrum.
}
\begin{tabular}{cllllllc}
\input{flux_stat_tab.2.dat}
\end{tabular}
\end{table}

\subsection{Mean Amount of Absorption}
\label{flxdec}

\begin{figure}
\includegraphics[width=84mm]{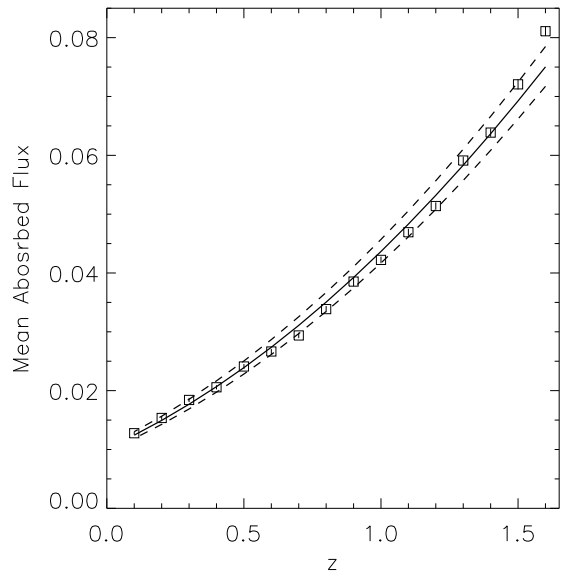}
\caption{The mean absorption as a function of redshift. The solid line
shows the fit through the points obtained from the optical depth fit
$F_{fit} = exp(-\tau_{eff}^{fit})$ in Fig. \ref{figtaueff}. The upper and
lower curves are fits to the 2$\sigma$ upper and lower
limits on each point.
}
\label{fluxdecrement}
\end{figure}

We define $DA$, the mean amount of absorption, as ${DA = 1 - \bar{F}}$,
where ${\bar{F}}$ is the mean flux in all the spectra that we made
at that $z$. In Fig. \ref{fluxdecrement}
we show the $DA$ as a function of redshift.  Since the
effective optical depth is $\tau_{eff} << 1$, we expect $DA = 1 -
\bar{F}=1-exp(-\tau _{eff}) \simeq \tau_{eff}$ as can be seen in the
figure.  The amount of absorption decreases by a factor of 6.2 from $z=1.6$
to 0.1. We know from \citet{jena05a} that had we used a simulation with a
cell size of 18~kpc instead of the 75~kpc used here,
the mean flux would be reduced by about 0.02 at $z=2$.

\subsection{Distribution of Flux}

\begin{figure}
  \includegraphics[width=84mm]{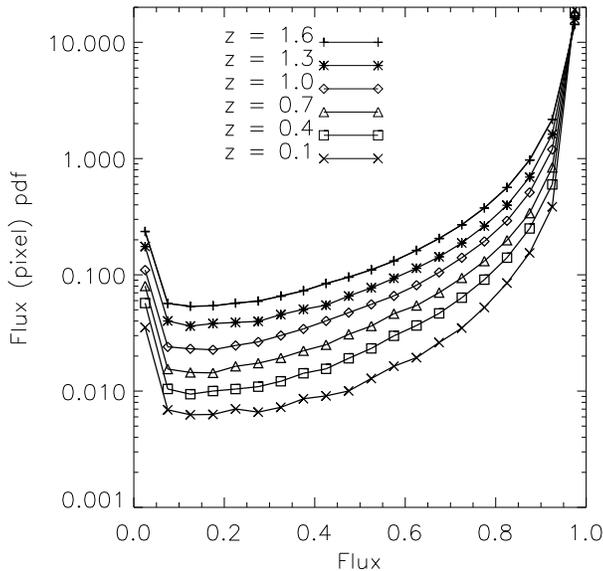}
  \caption{The pdf of the flux per pixel. We use flux bins of width
  0.05 from 0.0 -- 1.0, and we plot values at the bin centres, 0.025,
  0.075, ..., 0.975.
}
  \label{figpixelfluxpdf}
\end{figure}

In Fig. \ref{figpixelfluxpdf} we show the distribution of flux per pixel for
all spectra through the box, the flux pdf. The flux pdf is the number of
pixels in each range of flux, flux bins,  divided by the total number of pixels
and then divided by the bin width, which is 0.05. We have $1024$ pixels
per spectrum and $2000$ spectra per redshift.  Using this common
definition, the values unfortunately depend on the bin width. Our bins are
all the same width starting with the first bin at flux $0 < F < 0.05$
with a centre at 0.025, and a last bin at $0.95 < F < 1.0$. Our bin
centres and the widths of two of our bins differ from \citet{kim07b} and
\citet{bolton07d} who choose a first bin at flux $-0.025 < F <  0.025$ with
a centre at 0, and a last bin at $0.975 < F < 1.025$ with a centre at 1.0.
Since they divided all their bins by the same constant 0.05 (Kim private
communication), their flux pdf changes slope abruptly in the first
and last bins, being a factor of 2.0 below the value that we infer for
flux zero and 1.0. We do not follow their definition because it is
unexpected and hides trends.

In Fig. \ref{figpixelfluxpdf} we see a smooth and continuous change in the
shape of the flux PDf with redshift. As the redshift falls, the fraction
of pixels with $0.95 < F < 1.0$ increases while the fractions decline for
all other flux value. At $z=1.6$, 71\% of pixels have $F > 0.95$ and only
1.2\% have $F < 0.05$. At $z=0.1$, 95.6\% have $F > 0.95$ and 0.18\% have
$F < 0.05$. These values suggests that strong saturated lines that have
$F < 0.05$ are 6.7 times less common at $z=0.1$ compared to $z=1.6$.
This rate of change is similar to the factor of 6.5 decrease in the effective 
optical depth and 6.2 in the mean amount of absorption.

The fractional changes in the flux pdf values are similar for all
redshifts and fluxes, except for $F > 0.95$ where the curves for the
different redshifts cross. In detail, we see a faster relative change with
redshift at the lower fluxes; the curves spread out more to the left of
the plot. This suggests that strong deep lines that correspond to higher
column densities may evolve faster than do the weaker lines with smaller
column densities, a trend that we will see more clearly in other statistics.

\begin{table}
\caption{\label{pixel} The flux pdf at different redshifts. The first column is the
 redshift of the bin centre
 while the rest of the columns are the flux PDF values at different redshifts.
 All bins have a width of 0.05, starting with the first bin from flux 0 -- 0.05.
 The pdf values in the bulk of the table are the fraction of pixels with flux
 value in the bin, divided by 0.05. The flux pdf values in each column sum to 1/0.05.
}
\begin{tabular}{lcccccc}
\input{pixel_flux_pdf3.dat}
\end{tabular}
\end{table}

\subsection{Flux power}

\begin{figure}
\includegraphics[width=84mm]{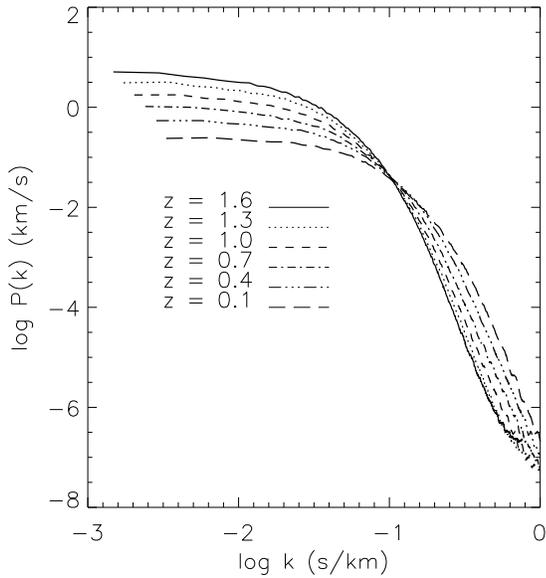}
\caption{The mean 1D flux power from all 2000 spectra at each redshift.
We list the values in Table \ref{Pfk}.  
}
  \label{figfluxpwr}
\end{figure}

We define the flux contrast as \begin{equation} f(u) = (F/\bar{F}) -1,
\label{fluxc} \end{equation} where $u$ is velocity and $\bar{F}$ is the mean
flux from the all spectra that we made at a given $z$.
We measure the one-dimensional
flux power of each sight line in the $z$ direction by computing the
fourier transform of the flux contrast, and we present the average of the
power from all sight lines at a given $z$ in Fig. \ref{figfluxpwr}.
The smallest $k$ value that we show is that for one wave along an
edge of the box. This is a constant in comoving Mpc, but in velocity
units it falls with the decrease in the Hubble constant at low redshifts.

\begin{table}
\caption{\label{Pfk} The flux power at different redshifts at selected values 
of the wavenumber $k$ (s/km). The first column lists the redshifts while the 
next four 
columns list the value of the logarithm of the flux power at the wavenumber 
listed on top of each column.
}
\begin{tabular}{ccccc}
\input{power_stat_tab.1.dat}
\end{tabular}
\end{table}

\begin{figure}
  \includegraphics[width=84mm]{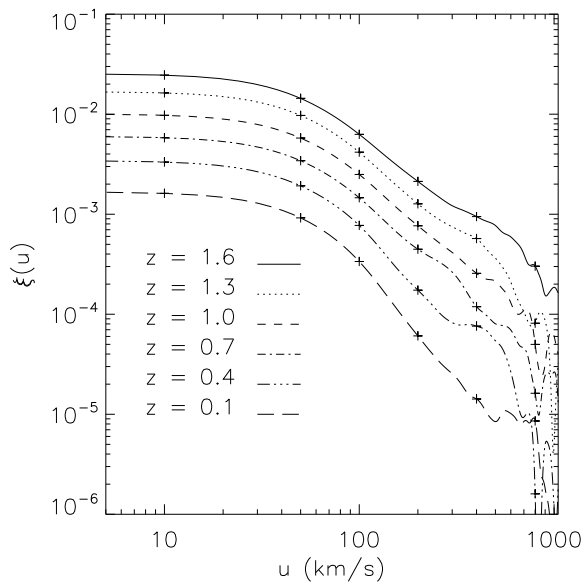}
 \caption{The flux autocorrelation function against the velocity lag.
 We use arbitrary curves to connect the points at a given redshift.
}
  \label{figfluxcor}
\end{figure}

We could compute the autocorrelation of the flux for a given velocity
lag $\delta u$ directly from a flux spectrum as $\xi_F(\delta u) =
<(F(u)-\bar{F}) (F(u+\delta u)-\bar{F}) >$, where $\bar{F}$ is the mean
flux for all spectra at that $z$ and the brackets refer to an average
across the pixels of the spectrum. Instead we choose to obtain the
autocorrelation from \begin{equation} \xi_F (\delta u) = { \bar{F}^2 \over
2\pi } \int P(k)e^{jk \delta u}dk \end{equation} where $P(k)$ is the power
of the flux contrast and we can use the line of sight averaged power
directly because we have divided by the mean flux of the spectra at that $z$.
In Fig.~\ref{figfluxcor} we show the average autocorrelation profiles at each
velocity lag from all lines of sight.

In Figs.  \ref{figfluxpwr} and \ref{figfluxcor} we again see
 clear systematic trends with $z$. As the redshift decreases the power
drops on large scales log $k < -1.0$ s/km, and increases on small scales.
The power
changes by less than a few percent at log $k = -1.0$ s/km (a wavelength of
62~\kms ) for $0.1 < z < 1.6$.
The decrease in power on larger scales at lower redshifts comes from
the decrease in the mean amount of absorption as the ionization of the IGM
increases.  The increase in the
power on small scale (large $k$) at low $z$ implies that the \lya\ lines
are narrower at low $z$, which we will see is correct in \S \ref{subsec:colden}.
As we saw in J05, had we used a cell size of about 18~kpc, the flux power at
$z=2$ would fall by about 15\% at $k=0.03$ and about 22\% at $k=0.01$ s/km.

The autocorrelation falls smoothly with both redshift and distance. It does
not show the crossing of curves that we saw for the power.

\subsection{Distribution of the mean flux per spectrum}

\begin{figure}
\includegraphics[width=84mm]{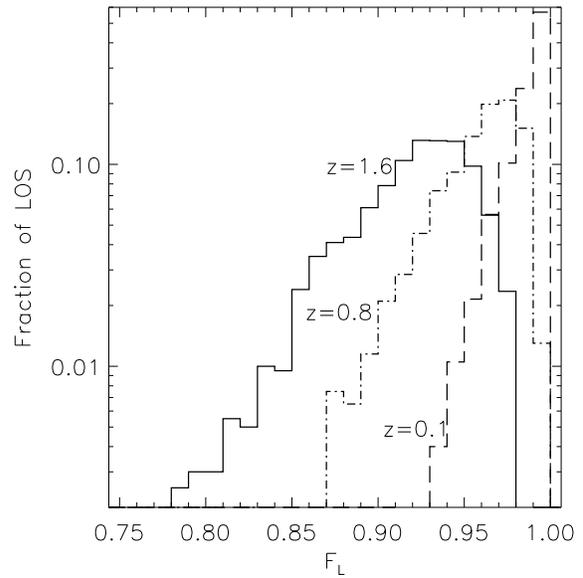}
\caption{The fraction of spectra, one per line of sight (LOS),
with a given mean flux $F_L$.
}
\label{figlinefluxpdf}
\end{figure}

In Fig. \ref{figlinefluxpdf} we show the distribution of mean flux per
line of sight.  We calculate the mean flux in each spectrum, and we plot
the fraction of the 2000 spectra with a given mean flux.
This statistic depends on the length of the spectra, which is
76~Mpc comoving. We would see less variation in the mean flux in
longer spectra.
As the redshift decreases the distributions shift to the right, to larger
mean fluxes, and the distributions become narrower, 
with less variation from spectrum to
spectrum. None of the spectra at $z=1.6$ has a mean flux $> 0.98$, where as
these are the most common mean fluxes at $z=0.1$.
These distributions, especially the tail containing the occasional
spectrum with a lot of absorption, reflect both the approximately
exponential distribution of \Lya\ line equivalent
widths and the clustering of matter on large scales.

\section{Lyman$\alpha$ Line Statistics }
\label{sec:linestat}

We obtain line statistics by fitting Voigt profiles as described in
\citet{zhang97}. As in \citet{tytler04b} and J05 (\S 5.1, 6.2)
for many functions and statistics we consider
only lines with $12.5 <$ \lnhi $< 14.5$~\cmm .

We shall examine two sample of lines; a full sample of all lines with
 central optical depths $\tau _o > 10^{-5}$ and a sub-sample with
 $\tau_o > 0.05$. A sample that was largely
complete to central optical depths $\tau_o > 0.05$
would require high enough resolution (FWHM $< 10$~\kms ) to completely
resolve all the narrowest lines that would otherwise appear too shallow.
The SNR per pixel must exceed 20 to
give $1\sigma $ in the centres of the limiting lines, barely enough to 
detect a line that is several pixels wide. To
detect most lines with $\tau _o = 0.05$ we would prefer SNR $> 60$.
The sample with $\tau_o > 10^{-5}$ includes a large number of additional shallow
very low column density ``lines" that are minor fluctuations
in the flux. Existing HST spectra have far too low SNR to show such features.

\subsection{Line widths as a function of column density }

\begin{figure*}
\includegraphics[width=84mm]{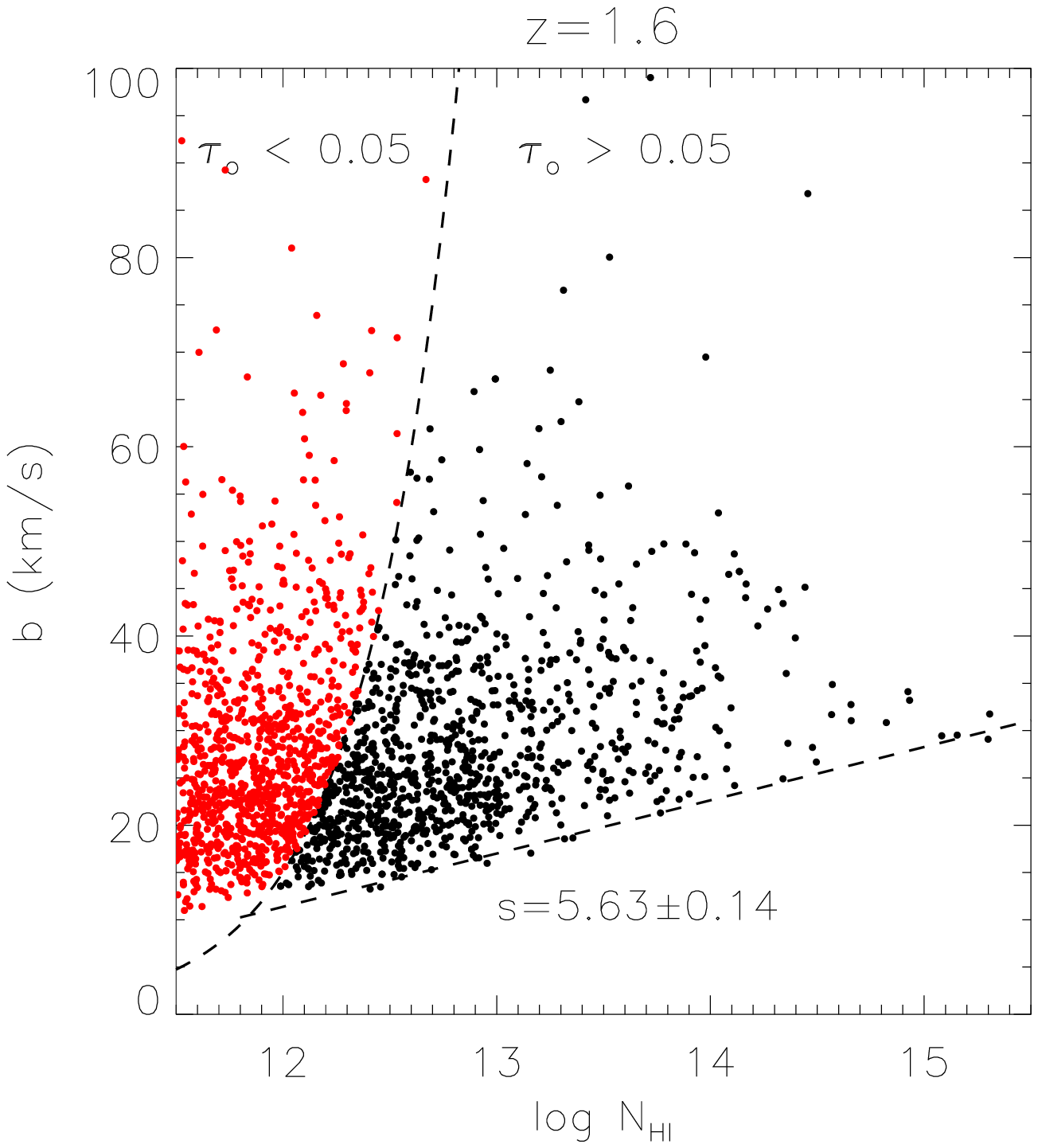}
\includegraphics[width=84mm]{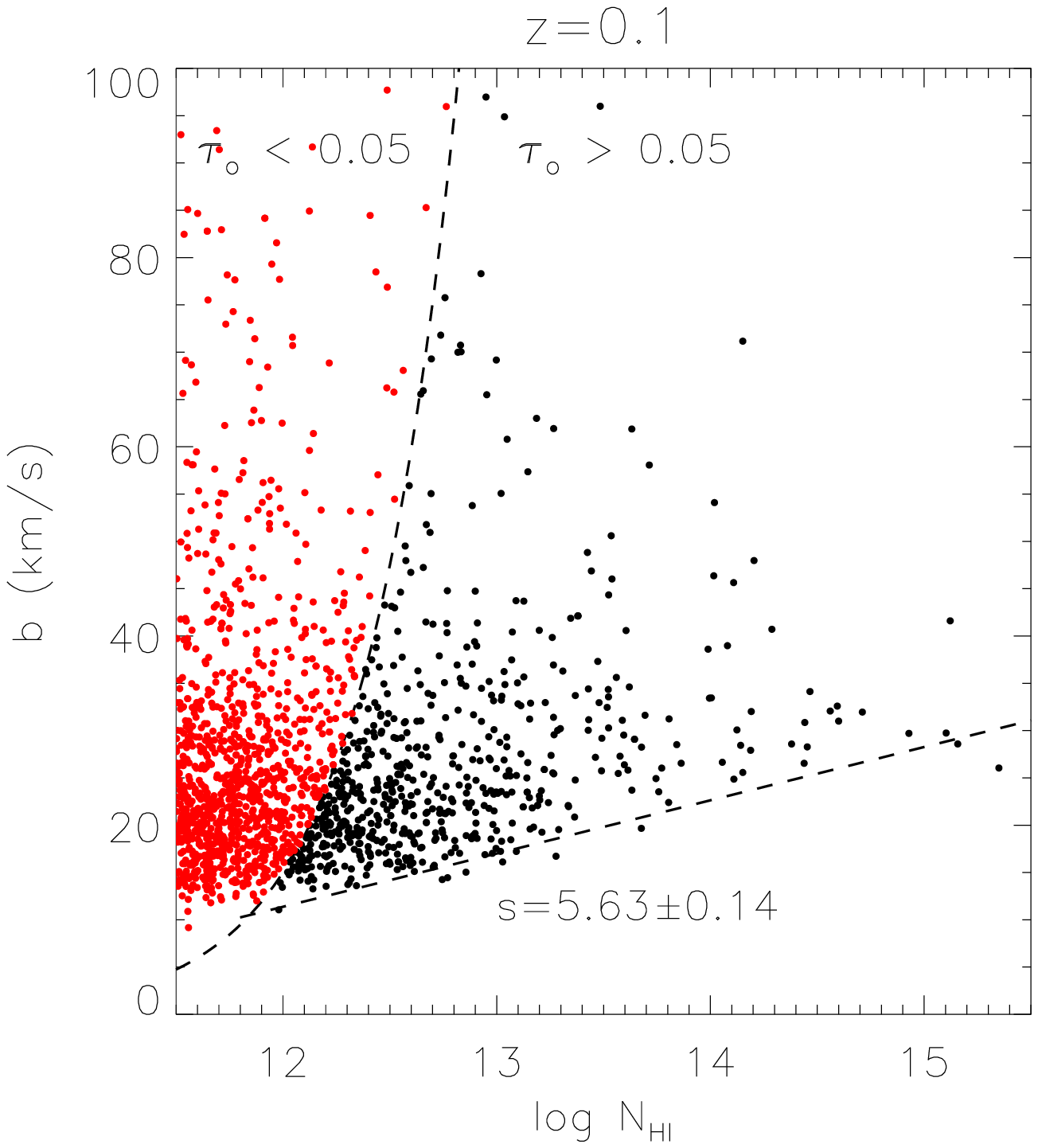}
\caption{Each point shows the $b$-value and \lnhi\ value (\cmm ) of a \lya\
line in one of the spectra at $z=1.6$ (left) and $z=0.1$ (right). We use 2000 
randomly
sampled spectra at z=1.6 and 1749 spectra at z=0.1. 
The relative number of points between the two redshifts corresponds to the 
ratio between the logarithm of the number of lines at z=1.6 (85817) and 
z=0.1 (20637) in the column density range of the plot, 
\lnhi $ = 11.5-15.5$~\cmm . 
The dashed curve shows lines with central optical
depth $\tau_{o} = 0.05$ from equation~(\ref{taucuteqn}).
Lines to the left of this curve  have
$\tau_{o} < 0.05$ and will be extremely hard to detect in HST spectra.
The dashed straight lines is a linear fit to the minimum $b$-values, 
$b_{min} = (5.63 \pm 0.14) log(N_{HI}/10^{13}) + (16.99 \pm 1.78)$ ~\kms\, 
This is a fit to the minimum in each of 10 bins
of width 0.2 from \lnhi $= 12.5-14.5$~\cmm\ at z=1.6.
We show the same straight line at z=0.1
}

\label{bnhscat}
\end{figure*}

In Fig. \ref{bnhscat} we show the  $b$-values that quantify
line width and H~I column
densities for \lya\ lines in spectra at redshifts 1.6 and 0.1.
We show only a fraction of all lines in our spectra at $z=1.6$ to avoid
saturating the plot.
We show a curve on the plots that marks lines with line centre optical
depth $\tau_{o} =0.05$.
From \citep{spitzer78},
\begin{equation}
 N_{HI} = \sqrt \pi \tau_{o} b / \sigma_o \lambda_o ,
\end{equation}
where $\lambda_o$ is the rest frame wavelength and $\sigma_o$ is the
cross-section. The curve has equation
\begin{equation}\label{taucuteqn}
b ~({\rm \kms }) = 15.0 (0.05/\tau_o) (N_{HI}/10^{12} \rm \cmm ),
\end{equation}
and appears concave on the log-linear plots.

We also show on the plots a straight line that represents the increase in
the minimum $b$-value as a function of \lnhi .
A visual examination of the plot shows
no major changes in the distribution in the $b$-$N$ plane with redshift,
but we will
see below that there are definite differences in the both the
$b$ and $N$ distributions.

\subsection{Column Density Distribution}
\label{subsec:colden}

In Table \ref{fntable} we list values for the H~I column density distribution,
$f(N)$, lines per unit $z$, per spectrum and per unit linear column density.
With this standard definition \citep{tytler87b}
the integral of $f(N)$ over a range of linear (not log) column
densities gives the number of lines in that column density
range per unit redshift:
\begin{equation}
\frac{dn(z)}{dz} = \int f(N) dN.
\end{equation}

\begin{table}
\caption{\label{fntable} The distribution of H~I column density values in
\Lya\ lines. We list values of log f($N_{HI}$) as a function of
\lnhi\ value in the first column and redshift given in the top row.
We include all lines with line centre optical depth $\tau _o > 10^{-5}$.
We use column density bins of width 0.2 for \lnhi = 12.0--16.0 \cmm ,
and 0.5 for \lnhi = 16.0--18.0~\cmm .
We list N/A when there are no lines in the column density bin
in any of our 2000 spectra at that $z$.
}
\begin{tabular}{lrrrrrr}
\input{fntable2.dat}
\end{tabular}
\end{table}

\begin{figure*}
\includegraphics[width=84mm]{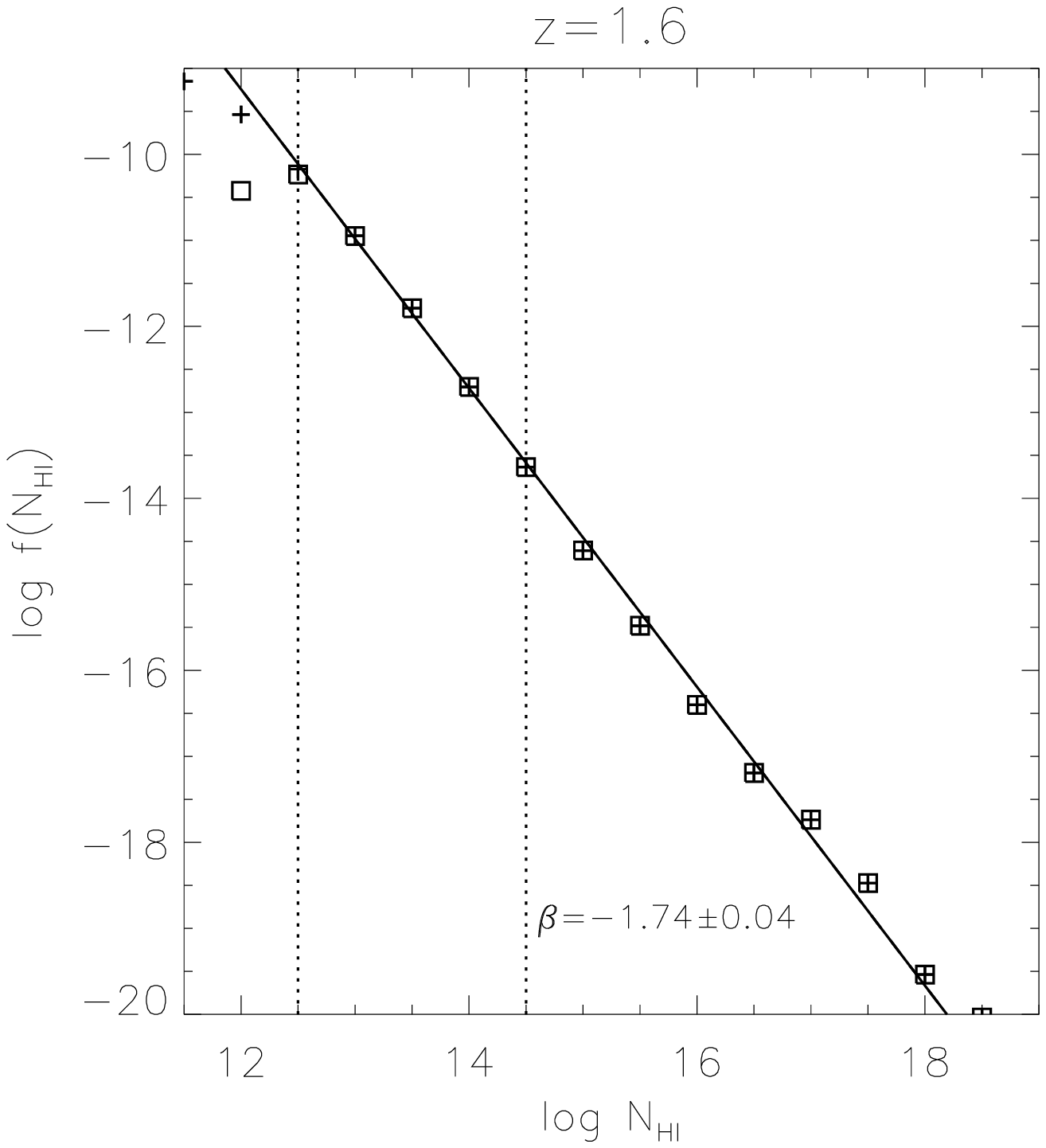}
\includegraphics[width=84mm]{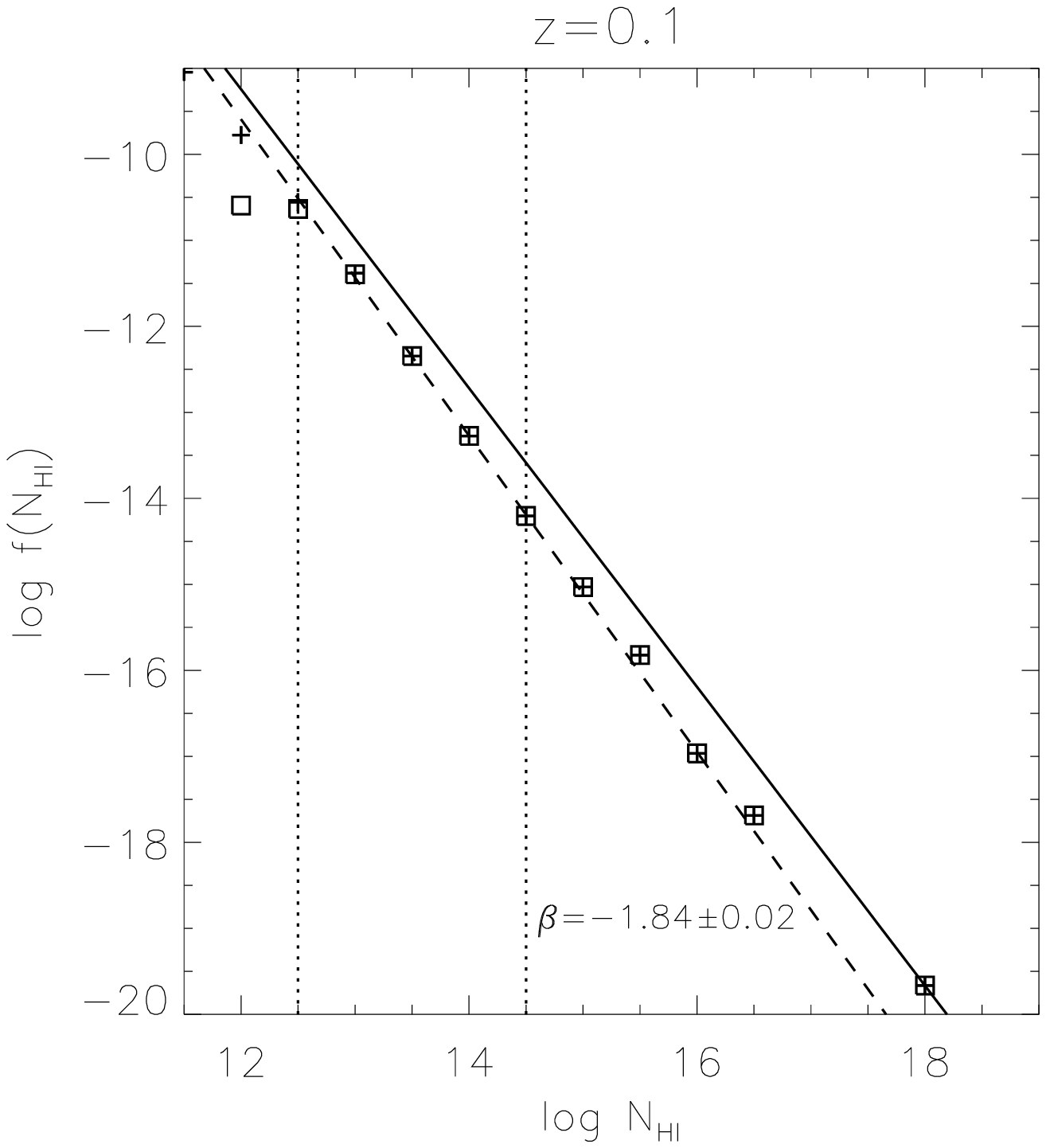}
\caption{The column density distribution $f(N)$
at $z=1.6$ and $z=0.1$.
The plus symbols show all lines with $\tau_{o} > 10^{-5}$, while the open squares
are lines with $\tau_{o} > 0.05$.
We show the same solid line on both plots. It is the power law,
$f(N_{HI}) \propto N_{HI}^{\beta}$ that best fits the number of lines with
$\tau_{o} > 10^{-5}$ and \lnhi $= 12.5-14.5$~\cmm\ in bins of 0.1, all at $z=1.6$.
The fit is $log f(N_{HI}) = (-1.74 \pm 0.04 ) log(N_{HI}/10^{12}) 
- (9.28  \pm 0.52 )$. We extend the line to all \lnhi\ values.
The dashed line is the power law fit at $z=0.1$,
$log f(N_{HI}) = (-1.84 \pm 0.02 ) log(N_{HI}/10^{12}) -
(9.61  \pm 0.34 )$, for the same column
density range.
}

\label{figcolumndensitydistribution}
\end{figure*}

In Fig. \ref{figcolumndensitydistribution} we show the $f(N)$ distribution.
A power law is a fair approximation to the  distribution at all the
$N$ values, although the plot scale obscures significant deviations.
We also show on the plot the sub-sample of the lines with $\tau _o > 0.05$.
This sub-sample contains all lines at all except the lowest columns
\lnhi $< 12.5$~\cmm .
On the plot of $f(N)$ for $z=0.1$ we show the fit to the $f(N)$ at
$z=1.6$ in addition to a fit to the $f(N)$ at $z=0.1$.
We see that there are fewer lines per unit $z$ at lower $z$ at all $N$ values,
and especially at high $N$ values. The slope of the $f(N)$ becomes
slightly steeper
at lower redshifts. As redshift drops, the number of lines with large
column densities falls faster than the number with small columns.

\subsection{The doppler parameter distribution}
\label{subsec:fb}

\begin{figure*}
\includegraphics[width=84mm]{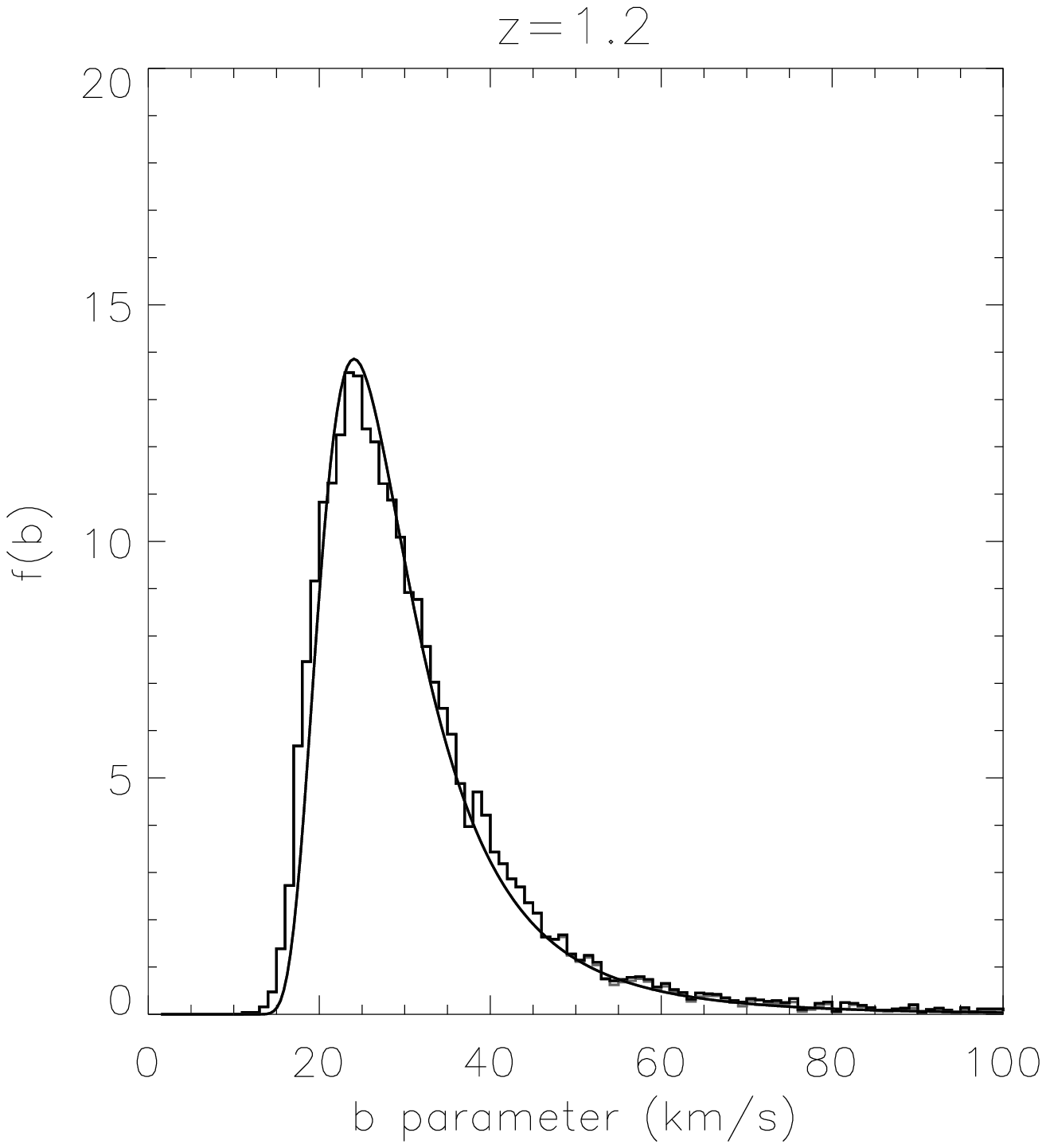}
\includegraphics[width=84mm]{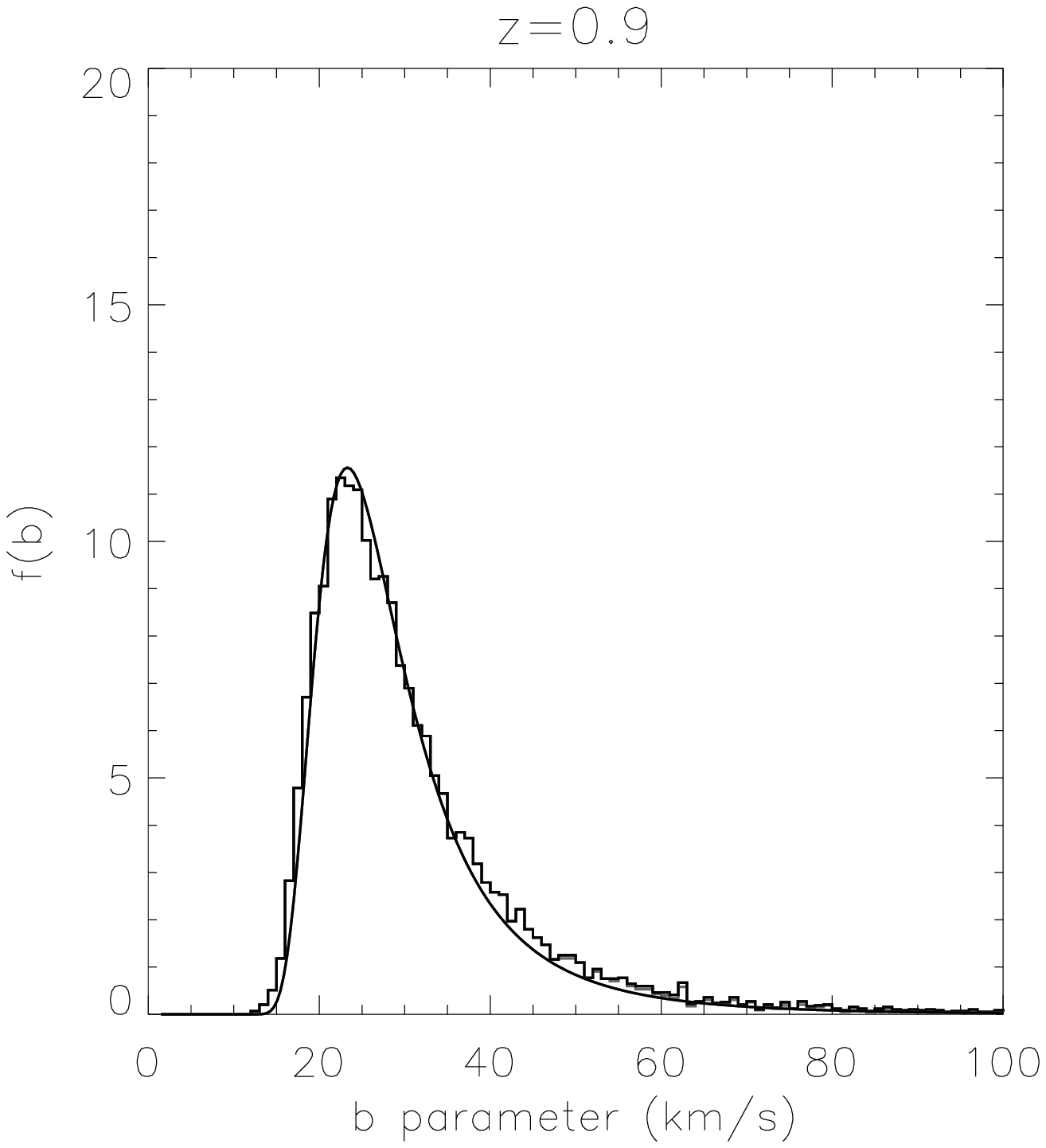}
\caption{
The distribution of the $b$-values $f(b) = d^2n/dzdb$ of \lya\ lines with
$log N_{HI}=12.5-14.5$~\cmm . The histogram shows the values in 1~\kms\ wide 
bins while the curve shows the Hui-Rutledge function.
The area under the histograms and curves is the number lines per unit redshift
with $12.5 < $ \lnhi\ $< 14.5$~\cmm .
}
\label{figd2ndzdb}
\end{figure*}

In Fig. \ref{figd2ndzdb} we show distribution of lines widths for redshifts
1.2 and 1.6, together with the Hui-Rutlidge \citep{hui99c} function,
\begin{equation}
\frac{dn}{db} = B\frac{b_{\sigma}^4}{b^5} exp(-\frac{b_{\sigma}^4}{b^4})
\end{equation}
where $dn/db$ is the number of lines per km s$^{-1}$ per unit $z$.
The Hui-Rutledge
function has only one parameter, \bsig , that describes the typical line
width, where the peak of the distribution is
 $b_{peak} = 0.9457 b_{\sigma}$.
In \citet[Fig. 18]{tytler04b} we showed that this function gave
an excellent fit to the distribution of $b$-values at $z=1.95$.
The function is also an excellent representation of the data at all low 
redshifts.
In detail, the data have a wider distribution than the function, with more
more lines at both small and large $b$-values.

\begin{figure}
\includegraphics[width=84mm]{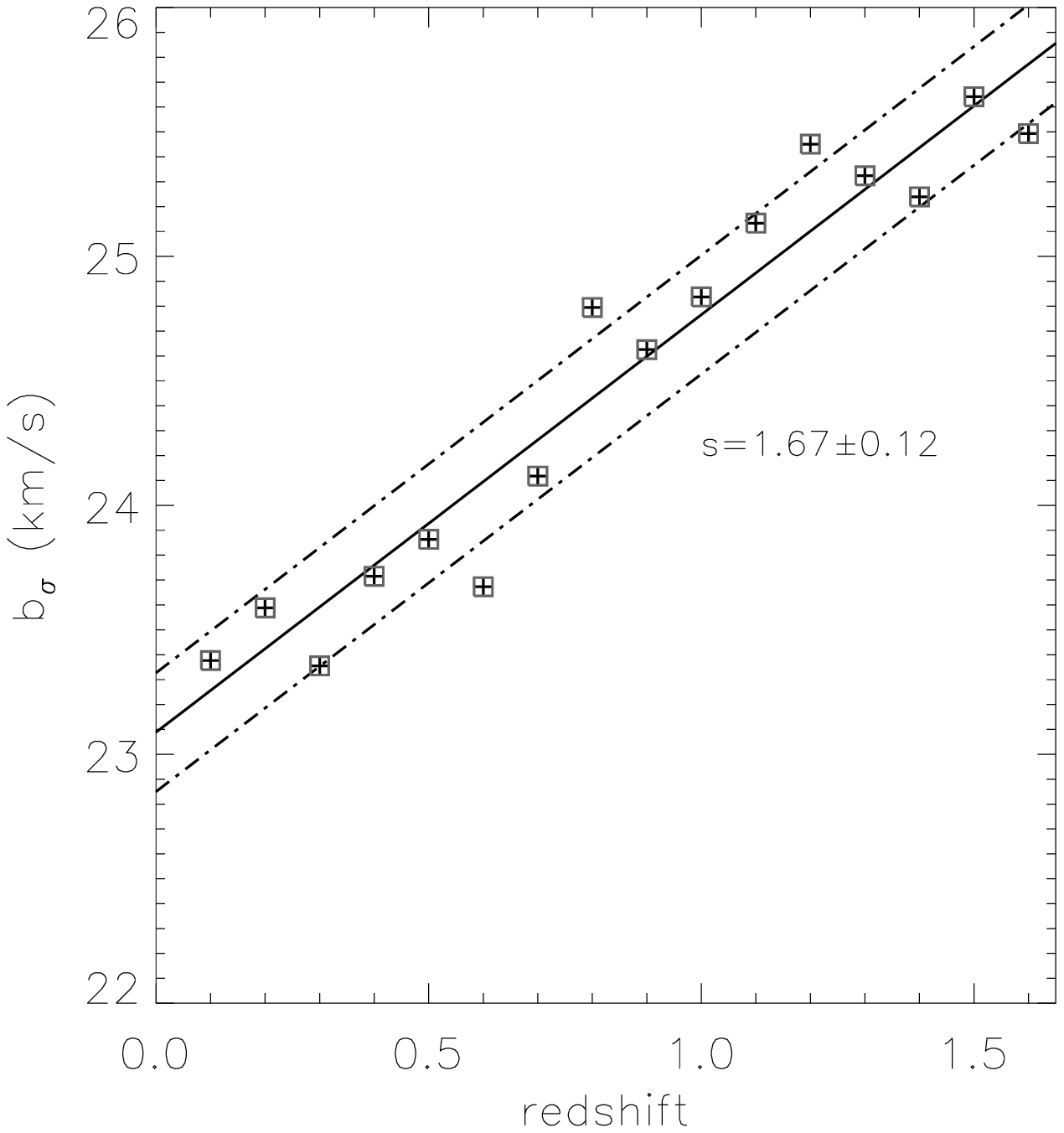}
\caption{The redshift evolution
of $b_{\sigma}$ from Table \ref{bdntable} from the simulated  spectra.
The solid line is a linear fit in the form 
$b_{\sigma} = (1.67 \pm 0.12)* z + (23.09 \pm 0.11) $ ~\kms\ .
The curves above and below the solid
line are best fits to the 2$\sigma$ upper and lower limits on the
individual points. They do not show the range of possible fits.
}
\label{figbsigbmed}
\end{figure}

We show in Fig. \ref{figbsigbmed} the $b_{\sigma}$ parameter as a
function of redshift. As previously, we show two different samples of lines;
all lines with $\tau _o > 10^{-5}$ and the sub-sample with
$\tau _ o > 0.05$.
  In both cases, the \bsig\ falls smoothly with redshift from 
$\sim 26.5$~\kms\ at $z = 1.6$ to $\sim 24.5$~\kms\ at
$z = 0.1$. The two samples give identical \bsig\ values because there are
very few lines with $\tau _ o < 0.05$ and \lnhi $> 12.5$~\cmm , the minimum
that we consider when we evaluate $f(b)$.
We know from \citet{jena05a} that a simulation with a cell size of 18~\kms\
 would give $b$-values smaller by approximately  1.0\%.

\begin{figure}
  \includegraphics[width=84mm]{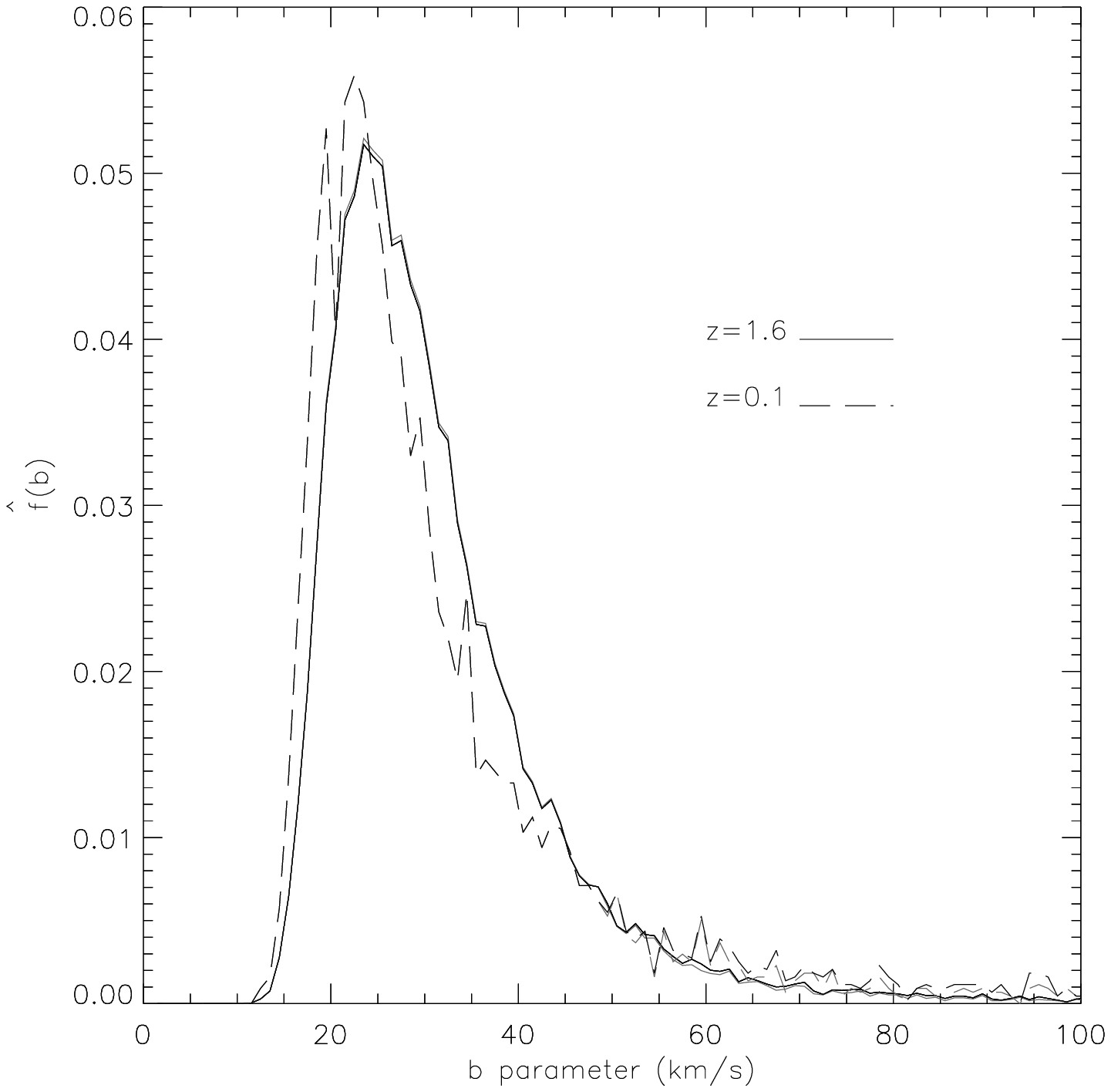}
  \caption{The doppler parameter distribution at two different redshifts. The 
distribution functions both have unit area.
}
  \label{figd2ndzdbnorm}
\end{figure}

In Fig. \ref{figd2ndzdbnorm} we show the $f(b)$ for redshifts 0.1 and 
1.6, each divided by  the number of lines per unit
redshift with column densities \lnhi = 12.5--14.5~\cmm , as listed in
listed in Table \ref{bdntable}. The distributions have very similar shapes,
although there are clearly more lines with smaller $b$-value at low redshift.

We list in Table \ref{bdntable} the $b_\sigma$ and the median $b$-value
$b_{med}$ (50\% of values are smaller) as a function of redshift. We prefer 
 the \bsig\ with high SNR spectra because this statistic isolates the most
 common line width and it is insensitive to
the precise shape of the $b$-value distribution at low and high values.
However in low SNR spectra, such as those from HST, the $b_{med}$ may be 
advantageous because it is even less sensitive to the detailed shape of the
distribution. In low SNR data, we expect a false excess of lines with low 
$b$-values and a lack of lines with large $b$-values.

\begin{table}
\caption{
\label{bdntable} Line widths as a function of redshift in simulated spectra. 
We list the Hui-Rutledge $b_{\sigma}$ parameter and the median $b$-value 
$b_{med}$ from lines with $\tau _o > 0.05$. The
last column contains the log of the number of lines per unit $z$ with
\lnhi = 12.5--14.5~\cmm .
}
\begin{tabular}{cccc}
\input{btable.dat}

\end{tabular}
\end{table}

\subsection{The equivalent width distribution}

\begin{figure}
\includegraphics[width=84mm]{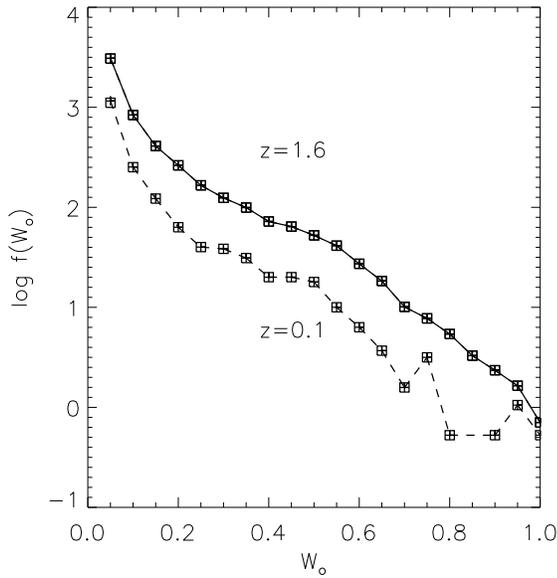}
\caption{The equivalent width distribution, defined as
$f(W_{o})=d^2n/dzdW_{o}$, shown at $z=1.6$ and $z=0.1$. $W_{o}$
is the rest frame equivalent width of an absorption line in
units of \AA~. The distribution is computed in bins of size 0.05 in
equivalent width and is shown for $W_{o} \geq 0.025$, the left edge of
the first bin which is cantered at $W_{o} = 0.05$.
The solid and dashed lines refer to $z=1.6$ and $z=0.1$ respectively.
Compared to data at $z=2$, our simulation lacks \lya\ lines with 
\lnhi\ $> 17$~\cmm\ or $W_o > 0.5$~\AA .
}
\label{figd2ndwdz}
\end{figure}

Fig. \ref{figd2ndwdz} shows the  distribution of the rest frame equivalent 
widths
$W_o$ for all lines at two redshifts. We list values in Table \ref{equivtable}.

\begin{table}
\caption{\label{equivtable} The equivalent width distribution, $f(W_o)$, 
with $W_o$ in ~\AA , at
different redshifts. A value of N/A means there were no lines at that 
equivalent width for that redshift.
}
\begin{tabular}{lllllll}
\input{equiv_width_dist.dat}
\end{tabular}
\end{table}

\begin{figure}
\includegraphics[width=84mm]{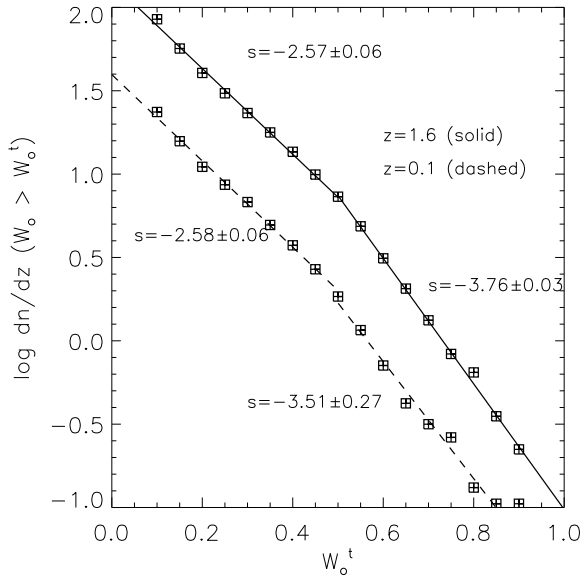}
\caption{The number of lines per unit $z$
with equivalent width exceeding $W_o^t$ (\AA ngstroms).
We show the results for $z=1.6$ (above) and $z=0.1$ (below). The straight lines
with slopes $s$ join at $W_o^t = 0.5$. Compared to data at
$z=2$ our simulation lacks \lya\ lines with $W_o > 0.5$~\AA .
}
\label{figdndzwc}
\end{figure}

Fig. \ref{figdndzwc} shows for two redshifts the number of lines with 
$W_o > W_o^t$, some minimum equivalent  width. We find that a pair of straight 
lines can represent the points at each redshift, hence the distributions are 
approximately broken exponentials; $ln (dn/dz) \propto W_o^t$.  

\begin{figure}
\includegraphics[width=84mm]{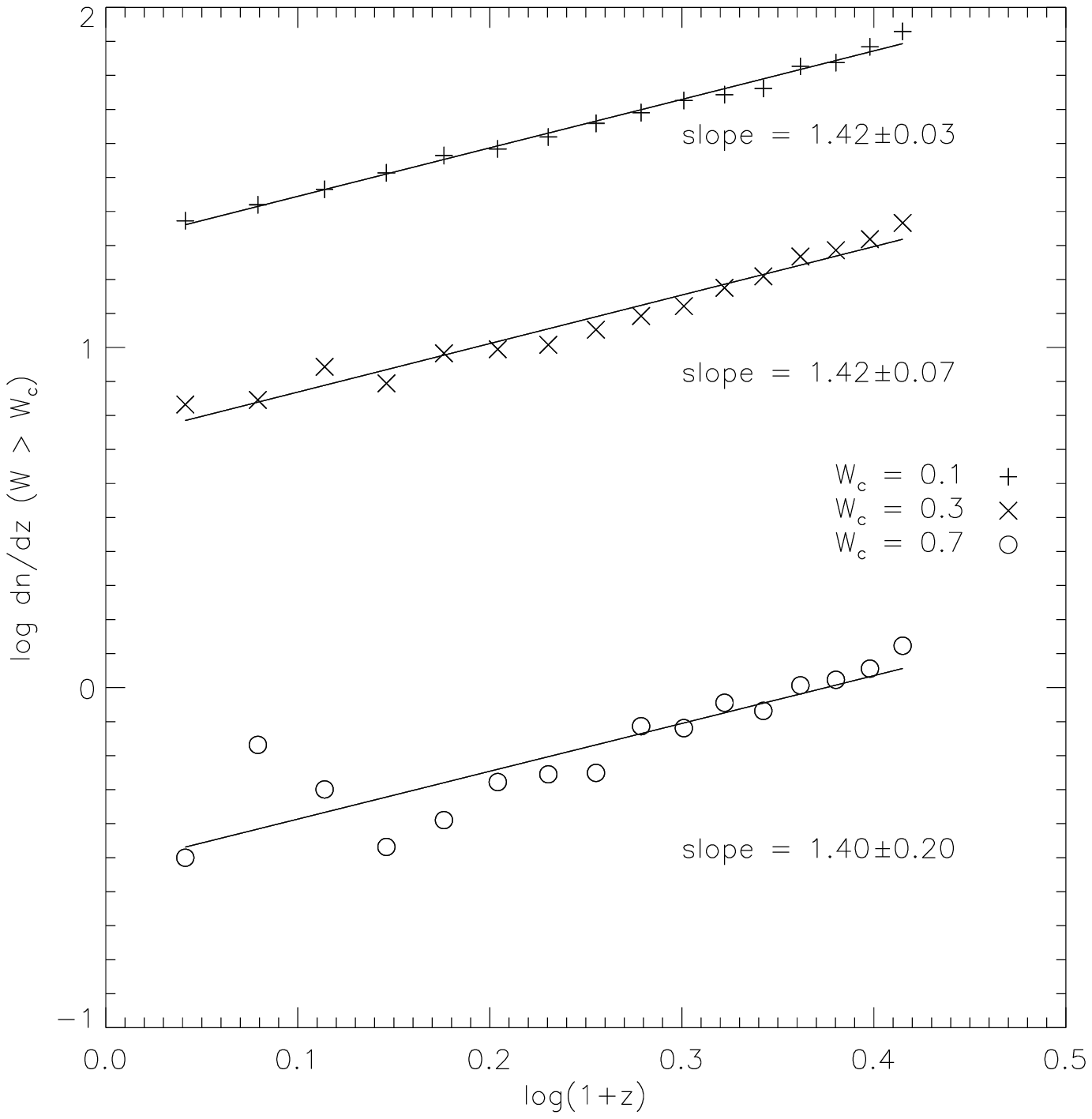}
\caption{
The redshift evolution of the number of lines
exceeding equivalent width $W_o$=0.1, 0.3 or 0.7 \AA ngstroms.
We also show power laws $dn/dz \propto (1+z)^{\gamma}$ with
slopes $\gamma $.
}
\label{figdndzwcslope}
\end{figure}

 In  Fig.~\ref{figdndzwcslope} we show how the number of lines with 
$W_o > W_o^t$ depends on redshift. We see that power laws are fair 
representations for all $z$ and $W_o^t$ values.

\begin{figure}
\includegraphics[width=84mm]{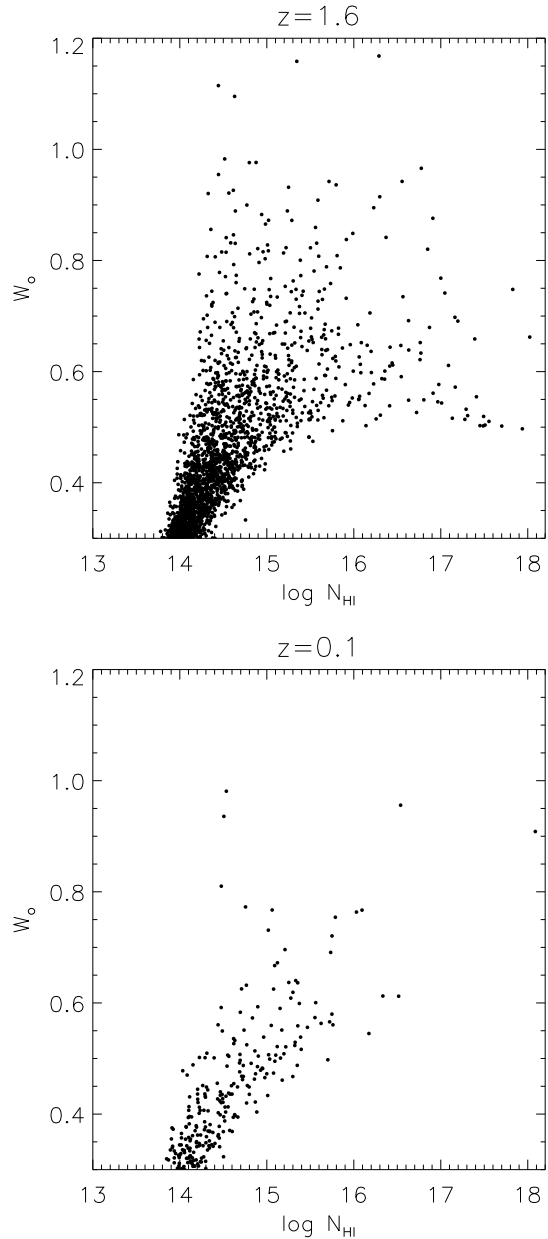}
\caption{The equivalent widths (\AA ngstroms) and H~I column densities (\cmm )
of \lya\ lines at $z=1.6$ and $z=0.1$. We show all lines in the plot range
from all 2000 
simulated spectra at each redshift, except for four with \lnhi $>18.2$~\cmm\
at $z=1.6$. 
}
\label{WvsNscatter}
\end{figure}

 In Fig. \ref{WvsNscatter} we show the equivalent widths as a function of
 column density. For low column densities, lines are unsaturated and on
 the linear portion of the curve of growth, where $W_o \propto $~\lnhi , there 
is no dependence on the $b$-value. These are the most common lines with 
 \lnhi\ $\sim 14$~\cmm\ and $W_o \sim 0.3$~\AA .
However, for larger \lnhi\ values, the lines become saturated and the
$W_o$ depends on  the $b$-value. We see that when $W_o > 0.5$~\AA\
we have some lines with  very large column densities. The simulation has
factors of several too few lines with \lnhi $> 17$~\cmm\ at $z=2$, and 
hence it also lack lines with $W_o > 0.5$~\AA . For saturated lines with a 
fixed
\lnhi\ value, we obtain a larger $W_o$ value when the $b$-value is larger
which allows more of the atoms to absorb photons because they are more widely
distributed in velocity. Hence the lines with $W_o \sim 1$~\AA\ and 
\lnhi $\sim 14.5$~\cmm\ have large $b$-values while those with $W_o \sim 0.5$~\AA\
and \lnhi $>16$~\cmm\ have small $b$-values. We know from 
\citet[Figs. 3 \& 5]{misawa04}
that at $z=2.5$ some such lines have $b$-values smaller than those of any lines 
with \lnhi $\sim 14$~\cmm. Note that the total $W_o$ from all lines 
in the simulated spectra with 
\lnhi $>17$~\cmm\ is negligible and hence these lines
 have a negligible effect on most statistics. We do not know 
how many lines we might see in a simulation with much smaller cell size and 
self-shielding.

\subsection{Number of Lines per unit redshift}

\begin{figure}
\includegraphics[width=84mm]{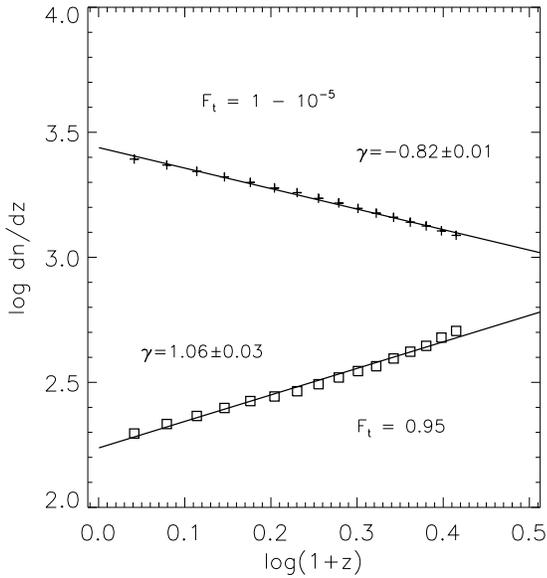}
\caption{The number of lines with central optical depth
$\tau_{o} > 10^{-5}$ (plus symbols) or
$\tau_{o} > 0.05 $ (open squares). We also show power
laws  $dn/dz =A (1+z)^{\gamma}$, with
${A = 3.439 \pm 0.004}$ and
 ${\gamma = -0.820 \pm 0.014}$ for $\tau_{o} > 10^{-5}$ (above)
 and
${A = 2.24 \pm 0.01}$
and ${\gamma = 1.06 \pm 0.03}$ for $\tau_{o} > 0.05 $ (below).
}
\label{figdndz}
\end{figure}

We show the number of lines per unit redshift in Fig. \ref{figdndz} for both the
large sample of lines with $\tau _ o > 10^{-5}$ and the sub-sample with
$\tau _ o > 0.05$. 
Note the dramatic difference! The large sample shows
a factor of a few more lines at lower redshifts, while the sub-sample, 
that is a better representation of data, shows the opposite. The number of 
lines
is clearly very sensitive to the definition of the sample \citep{zhang97},
yet in both cases the change with redshift is very well described by a 
single power law.

\begin{figure}
\includegraphics[width=84mm]{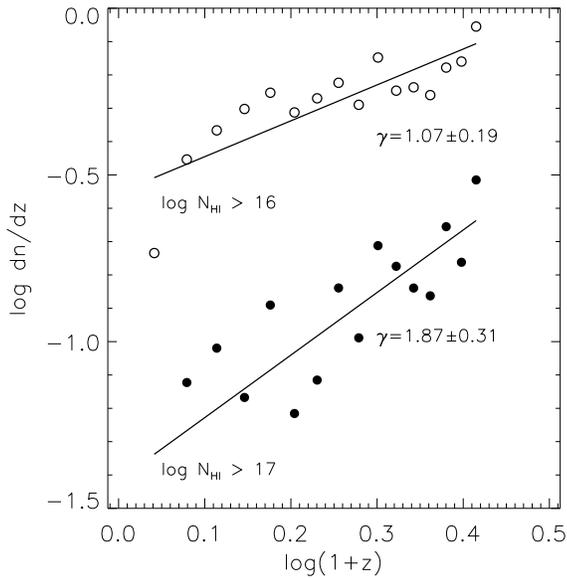}
\caption{
The redshift evolution of number of lines with
\lnhi $> 16$~\cmm\ (open circles) and \lnhi $> 17$~\cmm\
(filled circles). The straight lines are power laws
$dn/dz \propto (1+z)^{\gamma}$.
}
\label{dndzslope}
\end{figure}

In Fig.  \ref{dndzslope} we examine the redshift evolution of the
number of lines with very high column densities, \lnhi $> 16$ and \lnhi $>17$~\cmm .
The lines with \lnhi $> 17$~\cmm\ evolve more quickly.
We should remind the reader that our simulation has factors of many too few
lines with these column densities because it is optically thin and the cell 
size is factors of many too large.

\begin{figure}
\includegraphics[width=84mm]{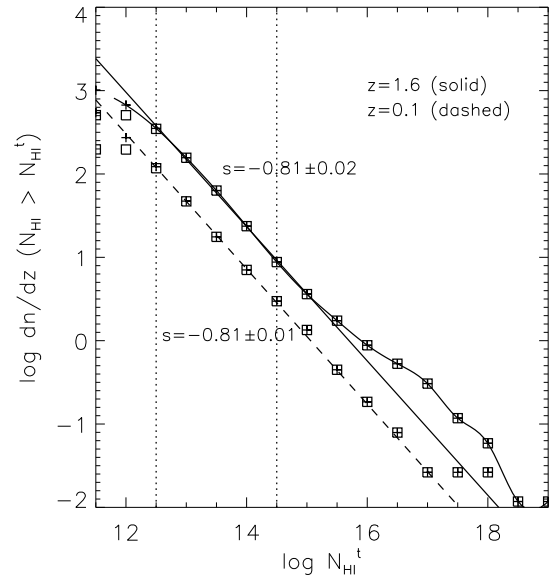}
\caption{The number of lines per unit redshift
as a function of the minimum column density
at $z=1.6$ and $z=0.1$. We show both the large sample of lines (plus symbols) and the 
sub-sample with $\tau_{o}>0.05$ (open squares). The solid straight 
line is a power law fit at $z=1.6$ (plus symbols) 
to \lnhi $^{t} = 12.5-14.5$~\cmm\ (vertical dotted lines).
The dashed straight line is the  corresponding fit to the $z=0.1$ points.
}
\label{dndzaboveNHI}
\end{figure}

We show in Fig. \ref{dndzaboveNHI} the number of \lya\ lines above a
cutoff column density at redshifts of 1.6 and 0.1. We see an excess
of lines above the power law fit at $z=1.6$
for \lnhi $ > 16$~\cmm . This excess is less at lower 
redshifts and it almost disappears by $z=0.1$.

\section{Comparison to Data}
\label{sec:datacomp}

Our simulations have known limitations. They lack high column density
lines compared to simulations that have much smaller cell size and
allow self-shielding. We also know that at $z=2$ there are differences between
 simulated and real spectra that show up in most statistics \citep{tytler07b}.
 
 When we compare to data we would ideally match the properties of the observed
spectra in detail, including the $z$ range, signal-to-noise ratio,
spectral resolution and continuum errors. In this section, we compare 
statistics of the \lyaf\ in real and simulated spectra.
 We will
ignore spectral resolution and the signal-to-noise ratio of the real spectra, 
although we will attempt to match the redshift and column density ranges.

\begin{figure}
  \includegraphics[width=94mm]{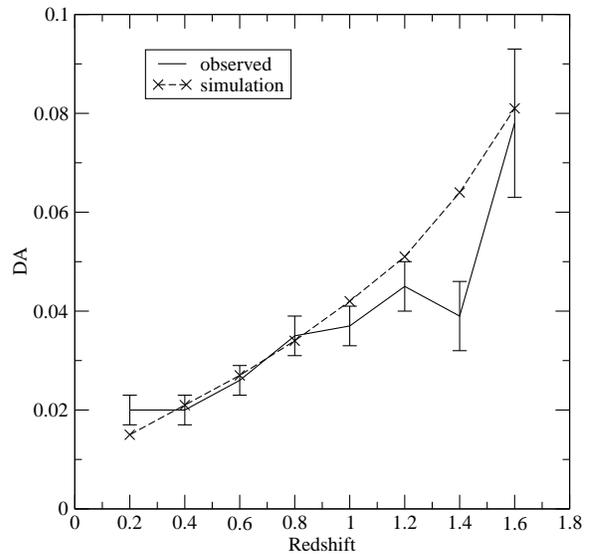}
  \caption{Comparison of the mean amount of absorption, DA values, in HST  spectra
  (error bars) from \citet{kirkman07a},
  and in the simulated spectra (crosses). We have not adjusted the
    the simulation output or spectra in any way.
}
  \label{dacomp}
\end{figure}

\citet{kirkman07a} measure the mean amount of absorption in the \lyaf\ at
low redshifts using $HST$ spectra of 74 QSOs.
In Fig.  \ref{dacomp} we show that our DA values are very
close to the observed ones from  \citet[Table 5]{kirkman07a}, where we 
do not consider the dip in DA at $z=1.4$ to be significant.

\citet{janknecht06a} fit the column densities and $b$-values to lines in
HST spectra of 9 QSOs at $z_{em} < 1.94$. They fit the column density
distribution with a power law ${f(N) = A \lnhi^{\beta}}$
and find no significant change of the power with redshift, just as we do.
They find a power law of $\beta \simeq 1.57 \pm 0.05$ consistent with
our value of $\sim 1.62$.

The median $b$-values from the HST spectra, 28~\kms ,  and from
our simulation,  28.5~\kms , are also very similar, see Table \ref{comptable}.
However, our \bsig\ values are larger than those in the HST spectra.
We have two comments on the \bsig\ values. It is well known that samples of 
lines
from spectra with significant noise, including all HST spectra, include
an excess of lines with low $b$-values. Common lines with the lowest column
densities enter the sample when photon noise gives them extra depth in
one or two pixels; see \citet[Fig. 10 \& 11]{kirkman97a} or 
\citet[Fig. 3]{misawa04}.
We saw  in J05 that the simulation also makes lines that are too wide at $z=2$ 
and that smaller cells would reduce the line widths but not enough to 
resolve the difference. We discussed this issue at length in \citet{tytler07b}.

\begin{table}
\caption{\label{comptable} Comparison of \lya\ lines in 
HST spectra \citep{janknecht06a} and in our simulated spectra. 
The HST spectra have
have 270 lines at $z = 0.5-1.0$ and 595 lines at $z = 1.0-1.5$.
$\beta $ is the index on the power law 
equation of the distribution of column densities, $f(N)$. 
The median line width is $b_{med}$ (\kms ) and \bsig\ is the Hui-Rutledge
parameter. 
}
\begin{tabular}{ccccccc}
\input{comptable.dat}
\end{tabular}
\end{table}

\section{Conclusions}
\label{sec:conc}

Some statistical descriptions of the \lyaf\ vary greatly over the
redshift range $z = 0.1 -1.6$  while others change little.
Statistics that change greatly derive from the mean amount of
absorption: the flux pdf, the power and auto correlation of the flux,
and the amplitude of $f(N)$ and the number of lines per unit $z$.
The statistics that change little relate to the line widths: the 
shape (not the amplitude) of the $b$-value distribution, $f(b)$,
the \bsig\ value and the $b$-$N$ distribution.

The changes in the \lyaf\ with redshift are all exceedingly smooth, 
with a constant rate of change across the redshift range 0.1 -- 1.6.
There are no special redshifts and no need for ``broken" power laws
or changes in slope. We can readily predict the values of parameters at 
larger or smaller redshifts. Specifically, we do not see the jump in
the evolution of the number of lines number of lines per unit $z$, 
as reported by \citet{weymann98}.
Rather, our results are very close to the constant rate of evolution
found in recent analysis of
larger samples of HST spectra by \citet{janknecht06a} and \citet{kirkman07a}.

We have found good agreement between the main statistics of the 
\lyaf\ lines in HST spectra and in our simulated spectra. These statistics
include the mean amount of absorption, the change in the absorption with
redshift and the median $b$-value. The \bsig\ is smaller in HST spectra than in
our simulations. This difference might be real, but given the
agreement in the median $b$-value, it is more likely
a result of the low SNR of the HST spectra giving a false excess of narrow 
lines.

The agreement between the \lyaf\ in our simulations and in HST spectra is 
better than we had anticipated. We already know that there are differences
between the simulation and data at $z=2$ but these are mostly too subtle to be
detected in the existing HST spectra. The HST spectra with high enough 
resolution to
resolve \lya\ lines are few and they have very low SNR by ground based 
standards.
While we had adjusted the astrophysical parameters of the simulation
to approximately match data at $z=2$ we made no such adjustments for any 
lower redshifts. The evolution of the \lyaf\ in the simulation is controlled 
by the
change in the gas density, the growth of structure and especially
the change in the ionisation and heating by the UVB. We know that the UVB 
intensity
was adjusted by Haardt and Madau to match a variety of data at
various redshifts, but they did not explicitly match the low redshift
IGM using a simulation of the \lyaf . 
They clearly did an excellent job of calculating the UVB at low redshifts. 
In this limited sense, our simulation is predicting the \lyaf\ at low redshifts,
and it is doing much better than we had expected.
This implies that the simulation comes close to representing the 
main physical process in the IGM, those responsible for the
appearance and evolution of the \lyaf .

\section*{Acknowledgments}

This work was supported by NASA  grant STScI AR-10688
and by NSF grant AST 0507717. We ran the simulation
using the facilities of the NCSA SDSC supercomputer centres with LRAC
allocation MCA98N020.

\bibliographystyle{mn2e}
\bibliography{ms9}

\clearpage

\end{document}

%% file: astmac.tex
\usepackage{color}
\usepackage{graphicx}
 
\newcommand{\Lya}{\mbox{Ly$\alpha$}}
\newcommand{\lya}{\mbox{Ly$\alpha$}}

\newcommand{\kms}{\mbox{km s$^{-1}$}}
\newcommand{\cmm}{\mbox{cm$^{-2}$}}

\newcommand{\sig}{\mbox{$\sigma_8$}}
\newcommand{\gammahe}{\mbox{$X_{228}$}}
\newcommand{\gammah}{\mbox{$\gamma_{912}$}}

\newcommand{\nhi}{\mbox{N$_{\rm H I}$}}
\newcommand{\lnhi}{\mbox{log \nhi}}

\newcommand{\bsig}{\mbox{$b_{\sigma}$}}

\newcommand{\etal}{{\it et al.}}
\newcommand{\lyaf} {\lya\ forest}

\newif\ifdraftmodep
\draftmodepfalse

\newif\ifapjp
\apjpfalse


%% file: flux_stat_tab.2.dat
$z$& $\Gamma_{-12}$ & $\tau_{eff}$ & DA &  err   &$\sigma_{L}^2$& $\sigma^2$  &$F_{mode}$ \cr
  & 		   & 		  &    &($10^4$)& ($10^3$)     &($10^2$)     &           \cr 
\hline
1.6 & 1.261 & 0.085 & 0.081 &  8.7 & 1.5 & 3.5 & 0.925 \cr
1.5 & 1.215 & 0.075 & 0.072 &  9.4 & 1.8 & 3.0 & 0.945 \cr
1.4 & 1.157 & 0.066 & 0.064 &  7.5 & 1.1 & 2.6 & 0.955 \cr
1.3 & 1.088 & 0.061 & 0.059 &  7.1 & 1.0 & 2.5 & 0.955 \cr
1.2 & 1.009 & 0.053 & 0.051 &  6.7 & 0.9 & 2.1 & 0.965 \cr
1.1 & 0.921 & 0.048 & 0.047 &  6.3 & 0.8 & 1.9 & 0.975 \cr
1.0 & 0.828 & 0.043 & 0.042 &  5.9 & 0.7 & 1.6 & 0.975 \cr
0.9 & 0.731 & 0.039 & 0.039 &  8.1 & 1.3 & 1.6 & 0.975 \cr
0.8 & 0.634 & 0.034 & 0.034 &  5.4 & 0.6 & 1.4 & 0.975 \cr
0.7 & 0.540 & 0.029 & 0.029 &  4.7 & 0.4 & 1.1 & 0.985 \cr
0.6 & 0.451 & 0.027 & 0.027 &  4.6 & 0.4 & 1.1 & 0.985 \cr
0.5 & 0.369 & 0.024 & 0.024 &  4.3 & 0.4 & 0.9 & 0.985 \cr
0.4 & 0.296 & 0.021 & 0.021 &  3.7 & 0.3 & 0.8 & 0.985 \cr
0.3 & 0.234 & 0.019 & 0.018 &  3.6 & 0.3 & 0.7 & 0.995 \cr
0.2 & 0.181 & 0.016 & 0.015 &  3.2 & 0.2 & 0.6 & 0.995 \cr
0.1 & 0.138 & 0.013 & 0.013 &  2.6 & 0.1 & 0.5 & 0.995 \cr

%% file: pixel_flux_pdf3.dat
flux  &           &           &  redshift &         &          &           \cr
      &       1.6 &       1.3 &       1.0 &       0.7 &      0.4 &       0.1 \cr
\hline
0.025 &     0.236 &     0.175 &     0.110 &     0.080 &    0.057 &    0.0354 \cr
0.075 &     0.057 &     0.040 &     0.024 &     0.015 &    0.011 &    0.0069 \cr
0.125 &     0.054 &     0.036 &     0.023 &     0.014 &    0.009 &    0.0062 \cr
0.175 &     0.054 &     0.038 &     0.023 &     0.014 &    0.010 &    0.0063 \cr
0.225 &     0.057 &     0.039 &     0.025 &     0.016 &    0.010 &    0.0070 \cr
0.275 &     0.059 &     0.040 &     0.026 &     0.017 &    0.011 &    0.0066 \cr
0.325 &     0.066 &     0.045 &     0.030 &     0.019 &    0.012 &    0.0072 \cr
0.375 &     0.073 &     0.050 &     0.034 &     0.022 &    0.014 &    0.0086 \cr
0.425 &     0.084 &     0.055 &     0.040 &     0.025 &    0.015 &    0.0091 \cr
0.475 &     0.095 &     0.066 &     0.047 &     0.031 &    0.019 &    0.0100 \cr
0.525 &     0.110 &     0.077 &     0.056 &     0.036 &    0.023 &    0.0128 \cr
0.575 &     0.131 &     0.093 &     0.066 &     0.046 &    0.030 &    0.0164 \cr
0.625 &     0.162 &     0.114 &     0.081 &     0.054 &    0.037 &    0.0194 \cr
0.675 &     0.207 &     0.143 &     0.105 &     0.070 &    0.047 &    0.0261 \cr
0.725 &     0.269 &     0.188 &     0.140 &     0.094 &    0.063 &    0.0348 \cr
0.775 &     0.376 &     0.263 &     0.193 &     0.131 &    0.091 &    0.0523 \cr
0.825 &     0.563 &     0.396 &     0.291 &     0.198 &    0.140 &    0.0849 \cr
0.875 &     0.972 &     0.693 &     0.511 &     0.339 &    0.250 &    0.1543 \cr
0.925 &     2.174 &     1.620 &     1.197 &     0.838 &    0.598 &    0.3848 \cr
0.975 &    14.201 &    15.828 &    16.970 &    17.937 &   18.551 &   19.1107 \cr

%% file: power_stat_tab.1.dat
        $z$ &	         &     log$k$ &		    &	          \cr  
            & 	   -2.0  &      -1.5  &       -1.0  & 	   -0.5   \cr
\hline 
      1.600 &      0.502 &      0.050 &      -1.382 &      -4.711 \cr
      1.500 &      0.417 &      0.005 &      -1.395 &      -4.693 \cr
      1.400 &      0.375 &     -0.042 &      -1.398 &      -4.656 \cr
      1.300 &      0.331 &     -0.064 &      -1.386 &      -4.541 \cr
      1.200 &      0.251 &     -0.132 &      -1.397 &      -4.462 \cr
      1.100 &      0.208 &     -0.167 &      -1.393 &      -4.342 \cr
      1.000 &      0.119 &     -0.223 &      -1.375 &      -4.261 \cr
      0.900 &      0.069 &     -0.256 &      -1.360 &      -4.118 \cr
      0.800 &     -0.018 &     -0.310 &      -1.361 &      -4.043 \cr
      0.700 &     -0.088 &     -0.376 &      -1.363 &      -3.919 \cr
      0.600 &     -0.148 &     -0.396 &      -1.347 &      -3.791 \cr
      0.500 &     -0.206 &     -0.439 &      -1.348 &      -3.645 \cr
      0.400 &     -0.338 &     -0.532 &      -1.356 &      -3.537 \cr
      0.300 &     -0.367 &     -0.578 &      -1.362 &      -3.425 \cr
      0.200 &     -0.496 &     -0.692 &      -1.410 &      -3.346 \cr
      0.100 &     -0.641 &     -0.776 &      -1.413 &      -3.232 \cr

%% file: fntable2.dat
\lnhi\ & \multispan{6} redshift \cr 
\cmm\  & 1.6  &    1.3  &    1.0  &    0.7  &    0.4  &    0.1 \cr
\hline 
  12.0 &  -9.51 &  -9.53 &  -9.56 &  -9.62 &  -9.67 &  -9.78 \cr
  12.2 &  -9.74 &  -9.79 &  -9.85 &  -9.92 &  -9.99 &  -10.15 \cr
  12.4 &  -10.01 &  -10.07 &  -10.15 &  -10.23 &  -10.28 &  -10.37 \cr
  12.6 &  -10.29 &  -10.38 &  -10.47 &  -10.54 &  -10.61 &  -10.71 \cr
  12.8 &  -10.61 &  -10.71 &  -10.80 &  -10.90 &  -10.94 &  -11.02 \cr
  13.0 &  -10.92 &  -11.04 &  -11.11 &  -11.21 &  -11.26 &  -11.36 \cr
  13.2 &  -11.26 &  -11.37 &  -11.46 &  -11.55 &  -11.62 &  -11.77 \cr
  13.4 &  -11.61 &  -11.71 &  -11.81 &  -11.88 &  -11.95 &  -12.16 \cr
  13.6 &  -11.95 &  -12.06 &  -12.12 &  -12.24 &  -12.35 &  -12.51 \cr
  13.8 &  -12.32 &  -12.43 &  -12.52 &  -12.65 &  -12.80 &  -12.92 \cr
  14.0 &  -12.69 &  -12.79 &  -12.92 &  -13.04 &  -13.17 &  -13.30 \cr
  14.2 &  -13.08 &  -13.17 &  -13.36 &  -13.51 &  -13.63 &  -13.58 \cr
  14.4 &  -13.44 &  -13.60 &  -13.70 &  -13.84 &  -13.95 &  -14.05 \cr
  14.6 &  -13.81 &  -13.97 &  -14.14 &  -14.14 &  -14.33 &  -14.33 \cr
  14.8 &  -14.23 &  -14.30 &  -14.42 &  -14.65 &  -14.79 &  -14.70 \cr
  15.0 &  -14.61 &  -14.66 &  -14.83 &  -14.97 &  -14.88 &  -15.02 \cr
  15.2 &  -14.96 &  -15.08 &  -15.21 &  -15.20 &  -15.26 &  -15.33 \cr
  15.4 &  -15.39 &  -15.35 &  -15.56 &  -15.50 &  -15.51 &  -15.60 \cr
  15.6 &  -15.55 &  -15.55 &  -15.94 &  -15.81 &  -15.87 &  -16.37 \cr
  15.8 &  -16.10 &  -15.93 &  -16.26 &  -16.23 &  -16.16 &  -16.20 \cr
  16.0 &  -16.37 &  -16.27 &  -16.18 &  -16.78 &  -16.83 &  -16.94 \cr
  16.5 &  -17.19 &  -17.27 &  -17.26 &  -17.40 &  -17.38 &  -17.69 \cr
  17.0 &  -17.74 &  -18.17 &  -17.97 &  -17.85 &  -17.88 &   N/A   \cr
  17.5 &  -18.47 &  -18.54 &  -18.68 &  -19.30 &  -19.23 &   N/A   \cr
  18.0 &  -19.54 &  -19.95 &  -19.88 &  -19.80 &     N/A &  -19.66 \cr 

%% file: btable.dat
	z	& $b_{\sigma}$	& $b_{med}$  & log$(dn/dz)$ \cr
\hline 
     1.6  &   25.49  &   28.38 & 2.53 \cr 
     1.5  &   25.64  &   28.25 & 2.50 \cr
     1.4  &   25.24  &   28.16 & 2.46 \cr
     1.3  &   25.32  &   28.13 & 2.43 \cr
     1.2  &   25.45  &   28.03 & 2.40 \cr
     1.1  &   25.13  &   27.97 & 2.37 \cr
     1.0  &   24.84  &   27.82 & 2.34 \cr
     0.9  &   24.62  &   27.36 & 2.31 \cr
     0.8  &   24.79  &   27.33 & 2.29 \cr
     0.7  &   24.12  &   27.08 & 2.25 \cr
     0.6  &   23.67  &   27.04 & 2.23 \cr
     0.5  &   23.86  &   26.90 & 2.21 \cr
     0.4  &   23.71  &   26.75 & 2.19 \cr
     0.3  &   23.35  &   26.72 & 2.15 \cr
     0.2  &   23.59  &   26.65 & 2.11 \cr
     0.1  &   23.38  &   26.62 & 2.08 \cr

%% file: equiv_width_dist.dat
 $W_o$ & \multispan{6}  redshift     \cr 
       &    1.6  &    1.3  &    1.0  &    0.7  &    0.4  &    0.1  \cr
\hline 
  0.05 &   3.490 &   3.380 &   3.299 &   3.199 &   3.160 &   3.064 \cr 
  0.10 &   2.923 &   2.826 &   2.738 &   2.655 &   2.583 &   2.403 \cr 
  0.15 &   2.613 &   2.513 &   2.428 &   2.330 &   2.208 &   2.087 \cr 
  0.20 &   2.419 &   2.289 &   2.225 &   2.112 &   1.976 &   1.801 \cr 
  0.25 &   2.220 &   2.137 &   1.997 &   1.882 &   1.787 &   1.602 \cr 
  0.30 &   2.094 &   2.003 &   1.926 &   1.727 &   1.531 &   1.585 \cr 
  0.35 &   1.997 &   1.873 &   1.662 &   1.625 &   1.501 &   1.492 \cr 
  0.40 &   1.858 &   1.804 &   1.643 &   1.523 &   1.328 &   1.301 \cr 
  0.45 &   1.807 &   1.696 &   1.582 &   1.464 &   1.372 &   1.301 \cr 
  0.50 &   1.719 &   1.591 &   1.464 &   1.376 &   1.420 &   1.253 \cr 
  0.55 &   1.616 &   1.542 &   1.302 &   1.256 &   1.201 &   1.000 \cr 
  0.60 &   1.435 &   1.325 &   1.258 &   0.981 &   0.957 &   0.801 \cr 
  0.65 &   1.263 &   1.154 &   1.041 &   0.926 &   0.860 &   0.567 \cr 
  0.70 &   1.004 &   0.957 &   0.852 &   0.584 &   0.355 &   0.199 \cr 
  0.75 &   0.889 &   0.717 &   0.740 &   0.583 &   0.355 &   0.500 \cr 
  0.80 &   0.733 &   0.341 &   0.464 &   0.487 &   0.258 &  -0.278 \cr 
  0.85 &   0.517 &   0.216 &    N/A  &   0.062 &  -0.042 &    N/A  \cr 
  0.90 &   0.371 &   0.392 &  -0.189 &  -0.115 &  -0.343 &  -0.278 \cr 
  0.95 &   0.216 &   0.040 &  -0.013 &  -0.416 &  -0.343 &   0.023 \cr 
  1.00 &  -0.151 &   0.040 &    N/A  &    N/A  &    N/A  &  -0.278 \cr 

%% file: comptable.dat
z 	  & $\beta$ & $\beta$ & $b_{med}$ & $b_{med}$ & $b_{\sigma}$ & $b_{\sigma}$        \cr
  	  & obs     & sim     & obs	  & sim       & obs          & sim                 \cr 
\hline 
0.5 - 1.0 & $1.57 \pm 0.05$   & $1.62 $   & 29        & 28.3         & 18.9         & 25   \cr
1.0 - 1.5 & $1.58 \pm 0.04$   & $1.63 $   & 28        & 28.5         & 23.5         & 25.5 \cr  